\documentclass[fleqn,usenatbib]{mnras}

\usepackage{newtxtext,newtxmath}

\usepackage[T1]{fontenc}

\DeclareRobustCommand{\VAN}[3]{#2}
\let\VANthebibliography\thebibliography
\def\thebibliography{\DeclareRobustCommand{\VAN}[3]{##3}\VANthebibliography}

\usepackage{graphicx}
\usepackage{amsmath}
\usepackage{calligra}

\newcommand{\be}{\begin{equation}}
\newcommand{\ee}{\end{equation}}

\def\ba{\begin{eqnarray}}
\def\ea{\end{eqnarray}}

\def\msun{M_\odot}
\def\msol{M_\odot}

\def\he3{^3He}

\def\ltsima{$\; \buildrel < \over \sim \;$}
\def\simlt{\lower.5ex\hbox{\ltsima}}
\def\gtsima{$\; \buildrel > \over \sim \;$}
\def\simgt{\lower.5ex\hbox{\gtsima}}
\def\vel{{\rm v}}
\def\velvec{{\rm \mathbf v}}

\newcommand{\cp}{\citep}
\newcommand{\ct}{\citet}

 \renewcommand{\vec}[1]{\mathbf{#1}}

\title[A study of convective core overshooting as a function of stellar mass]{A study of convective core overshooting as a function of stellar mass based on two-dimensional hydrodynamical simulations}

\author[I. Baraffe et al.]{I. Baraffe,$^{1,2}$ \thanks{E-mail: i.baraffe@ex.ac.uk}
J. Clarke,$^{1}$
A. Morison,$^{1}$
D. G. Vlaykov,$^{1}$
T. Constantino,$^{1}$
T. Goffrey,$^{3}$
T. Guillet,$^{1}$
\newauthor
A. Le Saux$^{1,2}$
and J. Pratt$^{4}$
\\
$^{1}$University of Exeter, Physics and Astronomy, EX4 4QL Exeter, UK \\
$^{2}$\'Ecole Normale Sup\'erieure, Lyon, CRAL (UMR CNRS 5574), Universit\'e de Lyon, France \\
$^{3}$Centre for Fusion, Space and Astrophysics, Department of Physics, University of Warwick, Coventry, CV4 7AL, UK\\
$^{4}$Lawrence Livermore National Laboratory, 7000 East Ave, Livermore, CA 94550, USA
}

\date{Accepted XXX. Received YYY}
\pubyear{2022}

\begin{document}
\label{firstpage}
\pagerange{\pageref{firstpage}--\pageref{lastpage}}
\maketitle

\begin{abstract}
We perform two-dimensional numerical simulations of core convection for zero-age-main-sequence stars covering a mass range from 3 $\msol$ to 20 $\msol$. The simulations are performed with the fully compressible time-implicit code MUSIC. We study the efficiency of overshooting, which describes the ballistic process of convective flows crossing a convective boundary, as a function of stellar mass and luminosity. We also study the impact of artificially increasing the stellar luminosity for 3 $\msol$ models. The simulations cover hundreds to thousands of convective turnover timescales. Applying the framework of extreme plume events previously developed for convective envelopes, we  derive overshooting lengths as a function of stellar masses.  We find that the overshooting distance ($d_{\rm ov}$) scales with the stellar luminosity ($L$) and the convective core radius ($r_{\rm conv}$). We derive a scaling law  $d_{\rm ov} \propto L^{1/3} r_{\rm conv}^{1/2}$ which is implemented in a 1D stellar evolution code and the resulting stellar models are compared to observations.   The scaling predicts values for the overshooting distance that significantly increase with stellar mass, in qualitative agreement with observations.   Quantitatively, however, the predicted values are underestimated for masses $\simgt 10 \msol$.
  Our 2D simulations show the formation of a nearly-adiabatic layer just above the Schwarzschild boundary of the convective core, as exhibited in recent 3D simulations of convection.  The most luminous models show a growth in size with time of the nearly-adiabatic layer.   This growth seems to slow down as the upper edge of the nearly-adiabatic layer gets closer to the maximum overshooting length and as the simulation time exceeds the typical thermal diffusive timescale in the overshooting layer. 
\end{abstract}


\section{Introduction}
One of the major uncertainties in stellar evolution models is the treatment of mixing taking place at convective boundaries \cp[see][]{Stancliffe16}.  
Convective motions do not  abruptly stop at the classical Schwarzschild boundary, but extend beyond it and lead to  the process of convective boundary mixing (CBM). 
The complex dynamics resulting from  convective flows penetrating in stable layers drives the transport of chemical species and heat, strongly affecting the structure and the evolution of stars.  
The same complex dynamics can also drive  transport  of angular momentum, impacting the rotational evolution of stars, the generation of magnetic field in their interior and their magnetic activity. 
CBM affects the evolution of all stars that develop a convective envelope, core or shell. Its treatment  is one of the oldest unsolved problems of stellar structure and evolution theory \cp[][]{shaviv73}. This extra mixing  could significantly alter the size of a convective core, the lifetime of major burning phases or the surface chemistry over a wide range of stellar masses. It  can impact the entire evolution of massive stars ($M \simgt 8 \msun$), determining  their structure before core-collapse supernova explosion and thus affecting nucleosynthetic yields which are crucial for galactic evolution studies \cp[][]{arnett11}. There is ample observational evidence pointing towards the need for extra internal mixing to explain a wide range of observations, such as eclipsing binaries \cp[][]{claret16}, color-magnitude diagrams \cp[][]{rosenfield17}  or asteroseismology \cp[][]{bossini15}. \ct[][]{rosenfield17} illustrate the uncertainty due to the treatment of core overshooting on ages and on morphological changes in stellar evolution tracks, significantly impacting stellar population studies. An increasing number of observational studies also suggests an increase of convective boundary mixing efficiency with stellar mass, using eclipsing binaries \cp[see][and references therein]{claret19} or Hertzsprung-Russell diagrams of massive stars \cp[][]{castro14}. In a recent study, \ct[][]{johnston21} confirms that current stellar models with no or with little convective boundary mixing usually under-predict the mass of convective cores. While such comparisons between stellar models and observations cannot identify a mechanism responsible for mixing at the convective boundaries,  \ct[][]{johnston21}  concludes that a range of efficiencies for the mixing mechanism(s) should be used. In addition to CBM, additional mixing could be due to rotation \cp[][]{zahn92} or internal gravity waves \cp[][]{schatzman93}. The latter are connected to CBM as they  are excited at convective boundaries by turbulent convective motions \cp[][]{press81, goldreich90, lecoanet13} and penetrating flows \cp[][]{rieutord95, montalban00, pincon16}. 

CBM is a generic term that encompasses different processes, namely {\it penetration}, {\it overshooting} or {\it entrainment}. The first term describes motions that cross a convective boundary and alter the background in such a way that the location of the convective boundary, defined by the Schwarzschild or the Ledoux criterion, moves inward or outward, resulting in the extension of the convective region. Overshooting usually describes convective penetrative motions that do not alter the background but  can still result in more or less efficient mixing \cp[][]{zahn91}. In the literature,  the terms overshooting and penetration are often used interchangeably. These processes have been described in stellar evolution models by an overshooting distance $d_{\rm ov}$ and/or a diffusion coefficient which remains constant or exponentially decays over the overshooting length \cp[][]{freytag96}. These parameters are usually calibrated to fit observations. The temperature gradient in the overshooting region is either set to the radiative or to the adiabatic temperature gradient \cp[see for example][]{michielsen19}. 
The third term {\it entrainment} is used to characterise shear-induced turbulent motions at the interface between the convectively stable and unstable regions driven by convective penetrative motions (plumes or eddies).  Interfacial instabilities 
contribute to mixing fluids of different compositions and/or densities, eroding the convective boundary.  This one can then grow in time following an entrainment rate characterised by the bulk Richardson number \cp[][]{fernando91, strang01}. Entrainment rates based on hydrodynamical simulations performed in a stellar context \cp{meakin07,jones17, cristini19}
are also implemented in stellar evolution codes to describe the extension of convective cores and shells \cp[][]{staritsin13, scott21}. However,  as shown by \ct[][]{scott21}, adopting  entrainment rates derived from existing stellar hydrodynamical simulations to main sequence stellar models produces unrealistic growth of the convective cores. The parameters that control the entrainment rates need to be decreased by several orders of magnitude to reproduce observations, questioning the reliability of the quantitative rates derived from  existing numerical simulations and even the existence of an entrainment process for main sequence convective cores. 

Describing and isolating these different processes characterising  CBM and at play at convective boundaries can be difficult in numerical simulations. Downward flows (or plumes) crossing a convective boundary at the bottom of an envelope are clearly observed in numerical simulations \cp[see for example][]{baraffe21}. Ballistic  plume crossings may eventually lead to  a modification of the thermal background -- the so-called {\it penetration} process. But for such modification to be observed, simulations must be run over many thousands of convective turnover timescales, as theoretically expected and recently  demonstrated in simulations by  \ct[][]{anders22} based on 3D simulations of convection in a Cartesian box with idealised setups. In a numerical study of solar-like convective envelopes, \ct[][]{baraffe21} show that artificially boosting the luminosity of the stellar model by a factor 10$^4$ yields a significant modification of the thermal background below the convective boundary with an extension of the size of the layer characterised by the penetration of convective flows, which could lead to a growth of the convectively unstable zone down to deeper levels. Whether this growth stabilises or whether the convective boundary continues moving downward indefinitely is unclear. 
For the solar-like model with realistic stellar luminosity, a slight modification of the thermal background is also observed in the simulations of \ct[][]{baraffe21}, but they show no trend of an extension of the Schwarzschild convective boundary over the simulation time. 

Following the approach developed in \ct[][]{pratt17} for convective envelopes, the most vigorous plumes can be used to define a maximal overshooting length, which can be significantly deeper than the typical length reached by the bulk of the plumes \cp[][]{pratt17, baraffe21, vlaykov22}. 
Whether this ballistic process is also observed  for convective cores and can drive significant mixing is an open question. Arguments based on the dynamics of  convective motions and plumes suggest that mixing below a convective zone (e.g envelope overshooting) and above (e.g core overshooting) may indeed be different \cp[][]{andrassy13}. Simple arguments based on the kinetic energy of a plume with typical velocity and the restoring buoyancy force suggest very small overshooting lengths  for the cores of low and intermediate mass zero-age-main-sequence (ZAMS) stars \cp[][]{higl21}. But these estimates are based on typical velocities without considering possible extreme plume events. The situation could also be different for convective cores on the ZAMS and on the main-sequence respectively. Indeed, the building of a molecular weight gradient at the core boundary due to hydrogen burning in the core can hamper the lifting of heavier material by ballistic processes.
An entrainment process slowly eroding the convective boundary may thus dominate at some point over the ballistic process during the main sequence evolution, or both processes may coexist and contribute  to mixing. These questions are still unsettled.   
Existing numerical simulations of convective cores have mostly focussed on one single stellar mass model, rather than a range of stellar masses  \cp{meakin07, gilet13, rogers13, edelmann19, horst20, higl21}. Additionally, many of these works  enhance the stellar luminosity of the model, to provide numerical stability, or to accelerate the thermal relaxation or the Mach number of the convective flow.
This artefact  may artificially favour one process over the other.  At this time, it is difficult to draw any firm conclusion regarding the main mechanisms that drive CBM in stars and how their efficiency is affected with stellar mass and with the stage of evolution on the main sequence.

In this work devoted to convective cores, we study the efficiency for convective plumes to penetrate into the stable region as a function of stellar mass for ZAMS models. In the following we will refer to overshooting to describe this process, since we essentially describe the ballistic process and even if a modification of the temperature gradient is observed for the most luminous models (see Sect. \ref{thermal}), likely leading to {\it penetration} as defined by \ct{zahn91}.  
We perform two-dimensional (2D) numerical simulations of convective cores of ZAMS stellar models covering a range of stellar masses between 3 $\msol$ and 20 $\msol$ (Sect. \ref{simulations}).  Our goal is to
apply the framework of extreme plume events developed for convective stellar envelopes \cp[][]{pratt17, pratt20, baraffe21} to the convective cores of intermediate and massive stars. We analyse whether extreme events can provide overshooting lengths required for stellar models to reproduce observations. 
For this purpose, we derive a relationship between overshooting length and stellar luminosity based on present numerical simulations (Sect. \ref{statistics}). We apply the relationship to one-dimensional stellar evolution models and test them against observations (Sect. \ref{1D}). This is the first step for a systematic study devoted to convective core overshooting in intermediate mass  and massive stars. 
 
\section{Numerical simulations}
\label{simulations}
We use the fully compressible time-implicit code MUSIC. A full description of MUSIC and of the time-implicit integration can be found in \ct{viallet11, viallet16, goffrey17}.  MUSIC
solves the inviscid Euler equations in the presence of external gravity and
thermal diffusion:

\begin{eqnarray}
\frac{\partial \rho}{\partial t} &=& - \vec \nabla \cdot (\rho \vec \velvec),\\
\frac{\partial \rho \vec \velvec}{\partial t} &=& - \vec \nabla \cdot (\rho \vec \velvec \otimes \vec \velvec)-\vec \nabla p + \rho \vec g,\\
\frac{\partial \rho e}{\partial t} &=& -\vec \nabla \cdot (\rho e \vec \velvec) - p \vec{ \nabla} \cdot \vec \velvec + \vec \nabla \cdot (\chi \vec \nabla T)
+ Q_{\rm nuc}, \label{eqeint}
\end{eqnarray}
\noindent where $\rho$ is the density, $e$ the specific internal energy, $\vec
\vel$ the velocity, $p$ the gas pressure, $T$ the temperature, $\vec g$ the
gravitational acceleration, and $\chi$ the thermal conductivity. The term  $Q_{\rm nuc}$ represents the nuclear energy rate.
The symbol $\otimes$ is the outer product. All  hydrodynamical simulations presented in this work are performed assuming spherically symmetric gravitational acceleration $\vec g$, which is updated every time interval $\Delta t$\footnote{Note that $\Delta t$  is  the time after which  the gravitational potential is updated, not the numerical timestep. The numerical timestep used for these simulations  is set by the hydrodynamical CFL number varying between 10 and 50 \cp[see][for definitions]{viallet11} and corresponding to values for the timestep  ranging between 5 s and 40 s.}.
All simulations presented in this work are performed with $\Delta t = 10^3$ s. The typical dynamical timescale of the entire stellar cores analysed in this study $\tau_{\rm dyn} \sim 1/\sqrt{(\rho_{\rm mean} G)}$, with $\rho_{\rm mean}$ the mean density of the core and $G$ the gravitational constant, is of the order of 10$^3$ s.  We have checked with a number of test simulations that a variation of $\Delta t$  between $10^2$ and $10^5$ seconds does not impact our results.
 
In the stellar models considered, radiative transfer is the major heat transport that contributes to the thermal conductivity, which is given for photons by
\begin{equation}
\label{eq:chirad}
\chi = \frac{16 \sigma T^3}{3\kappa \rho},
\end{equation}

\noindent where $\kappa$ is the Rosseland mean opacity, and $\sigma$ the Stefan-Boltzmann constant. 
Realistic stellar opacities and equation of states appropriate for the description of stellar interiors are implemented in MUSIC. Opacities are interpolated from the OPAL tables \cp{Iglesias96} for solar metallicity and the equation of state is based on the OPAL EOS tables of \ct{rogers02}.  

\begin{table*}
   \caption{Properties of the initial stellar models (all models have a central helium abundance $Y_{\rm c}$=0.2838) used for the 2D hydrodynamical simulations: total mass, stellar luminosity, stellar radius, mass and radius of the convective core (corresponding to the location of the Schwarzschild boundary) and the pressure scale height at the Schwarzschild boundary. }
   \label{tab1}
   \centering
   \begin{tabular}{c c c c c c } 
     \hline \hline
     $M/\msol$ &  $L_{\rm star}/L_\odot^{a}$ & $R_{\rm star}$ (cm) &  $M_{\rm conv}/\msol$ &  $r_{\rm conv}/R_{\rm star}$ & $H_{P,{\rm CB}}$ (cm) \\
      \hline
      3 &  7.7673 $\times$ 10$^{1}$& 1.3855 $\times$ 10$^{11}$ &  0.5724 & 0.1486 & 1.3 $\times$ 10$^{10}$ \\
      5 & 5.2186 $\times$ 10$^{2}$ & 1.8424 $\times$ 10$^{11}$ & 1.212 & 0.1814 & 1.8 $\times$ 10$^{10}$ \\
      10 & 5.5726 $\times$ 10$^{3}$  & 2.7295 $\times$ 10$^{11}$ & 3.046 & 0.2239 & 2.7 $\times$ 10$^{10}$ \\
       15 & 1.9242 $\times$ 10$^{4}$ & 3.4255 $\times$ 10$^{11}$ & 5.600 & 0.2580 & 3.3 $\times$ 10$^{10}$ \\
       20 & 4.2962 $\times$ 10$^{4}$ & 4.0172 $\times$ 10$^{11}$ & 8.7947 & 0.2869 & 3.7 $\times$ 10$^{10}$ \\
      \hline
      \multicolumn{3}{l}{$^a$ We use $L_\odot = 3.839 \times 10^{33}$ erg/s.}
   \end{tabular}
\end{table*}

\subsection{Initial stellar models}

To provide the initial structures for the 2D simulations,  we compute stellar models in the mass range 3-20 $\msun$ with the one-dimensional Lyon stellar evolution code \cp{baraffe91, baraffe98}, using the same opacities and equation of state as MUSIC\footnote{The 1D initial structures are available on the repository http://perso.ens-lyon.fr/isabelle.baraffe/2Dcore$\_$overshooting$\_$2023}. The 2D simulations require as initial input a radial profile of density and internal energy. The 1D stellar evolution models have an initial helium abundance in mass fraction $Y$=0.28 and solar metallicity $Z$=0.02 and were computed through the pre-main sequence and main sequence phases. All initial models for the 2D simulations in this study are taken at the beginning of core hydrogen burning and have a central abundance of helium $Y_{\rm c}$=0.2838,  {\it i.e.} only $\sim$ 1\% of their central hydrogen has been depleted. There is thus a very shallow mean molecular weight gradient at the convective boundary. Follow-up analysis  of later stages of evolution with a steeper gradient of molecular weight at the core boundary are in progress (Morison et al. in prep). Convective stability is defined by the Schwarzschild criterion $\nabla < \nabla_{\rm ad}$, with $\nabla = {{\rm d} \log T \over {\rm d} \log P}$ the temperature gradient and $\nabla_{\rm ad} = {{\rm d} \log T \over {\rm d} \log P}|_S$ the adiabatic gradient. The 1D stellar models used to generate the initial structures for the 2D simulations do not account for overshooting at the convective core boundary. In the following, we define the Schwarzschild boundary  as the transition layer between convective instability ($\nabla > \nabla_{\rm ad}$) and stability ($\nabla < \nabla_{\rm ad}$). The properties of the initial 1D stellar structures are provided in Table \ref{tab1}. 
Nuclear energy generated in the convective cores is accounted for in the internal energy equation (Eq. (\ref{eqeint})) through the term $Q_{\rm nuc}$  using the radial profile of the nuclear energy rate from the 1D stellar model. Given that the simulation times are orders of magnitude smaller than the nuclear timescale for H burning in the cores, the nuclear energy is assumed to remain constant with time.  

\subsection{Spherical-shell geometry and boundary conditions}
Two-dimensional simulations are performed in a spherical shell using spherical coordinates, namely  $r$ the radius and $\theta$ the polar angle, and assuming azimuthal symmetry in the $\phi$-direction. For all models, the inner radius $r_{\rm in}$ is defined at 0.02 $R_{\rm star}$. 
The choice of the outer radius  $r_{\rm out}$ depends on the stellar model. Since the main motivation of this work is to analyse the extent of the overshooting layer for different stellar masses, the outer radius $r_{\rm out}$ is fixed at a distance of $\sim 1 \times H_{P,{\rm CB}}$ for the lowest mass (3 $\msol$) to $\sim 3.5 \times H_{P,{\rm CB}}$ for the highest mass (20 $\msol$) away from the convective boundary $r_{\rm conv}$. Extension of the radial domain  to analyse the generation of internal waves at the core boundary
and their propagation in the radiative envelope is work in progress. 
The angular extent ranges from $\theta = 0 ^\circ$ to $\theta = 180 ^\circ$. The grid has uniform spacing in the r and $\theta$ coordinates. The choice for the resolution ($N_r, N_\theta$) is set by the condition to have a good resolution 
 of the pressure scale height at the Schwarzschild boundary.   
 Effective Reynolds and Prandtl numbers are commonly used to set the resolution of numerical simulations. But
 given that our simulations are based on an implicit Large Eddy Simulation (ILES)
approach, only  a rough estimate can be provided for these numbers. They will in
any case remain far away from the conditions prevailing in stellar interiors. We suggest that a more relevant
resolution criterion for hydrodynamical simulations devoted to the study of overshooting using
realistic stellar structures should be the number of grid cells per pressure scale height at the convective
boundary. This should allow a more relevant comparison between the works of different groups
devoted to the study of different stars. We use $\sim 110-140$ grid cells per pressure scale height in the radial direction. 
The details of the resolution adopted in this work are provided in Table \ref{tab2}. We have also performed a few tests with higher resolution and analyse the impact in Sect. \ref{statistics}.

The radial boundary conditions for the density correspond to a constant radial derivative on the density \cp[see][]{pratt16}. 
The energy flux at the inner and outer radial boundaries are set to the value of the energy flux at that radius in the one-dimensional stellar evolution model. At the boundaries in $\theta$, because of the extension of the angular domain to the poles,  reflective boundary conditions for the density and energy are used ({\it i.e.} the values are mirrored at the boundary).
For the velocity, we impose reflective conditions at the radial and polar boundaries, corresponding to:
\begin{itemize} 
\item{}  $\vel_r$ = 0 and ${\partial \vel_\theta \over \partial r}=0$ at $r_{\rm in}$ and $r_{\rm out}$,
\item{}  ${\partial \vel_r \over \partial \theta}=0$ and  $\vel_\theta$ = 0 at $\theta=0^\circ$ and $\theta=180^\circ$.
\end{itemize}

We have also performed simulations for the 3 $\msun$ model with artificial enhancement of the stellar luminosity 
and the thermal diffusivity by factors 10, 10$^2$, 10$^3$ and 10$^4$. This covers the range of luminosities of the stellar masses considered in this work (3-20 $\msun$). This choice of enhancement factor allows a comparative analysis of the impact of the luminosity for fixed core mass and increasing core mass, respectively.
 Note that even larger enhancement factors (up to $10^7$) for a 3 $\msun$ stellar structure can be found in previous works \cp[e.g.][]{rogers13, edelmann19}. For the artificially boosted simulations, the energy flux  (equivalently the luminosity)  at the radial boundaries is multiplied by the enhancement factor, the nuclear energy rate is multiplied by the same factor and the Rosseland mean opacities $\kappa$ in MUSIC are decreased by the same factor.

 \begin{figure}
\includegraphics[height=11cm,width=9cm]{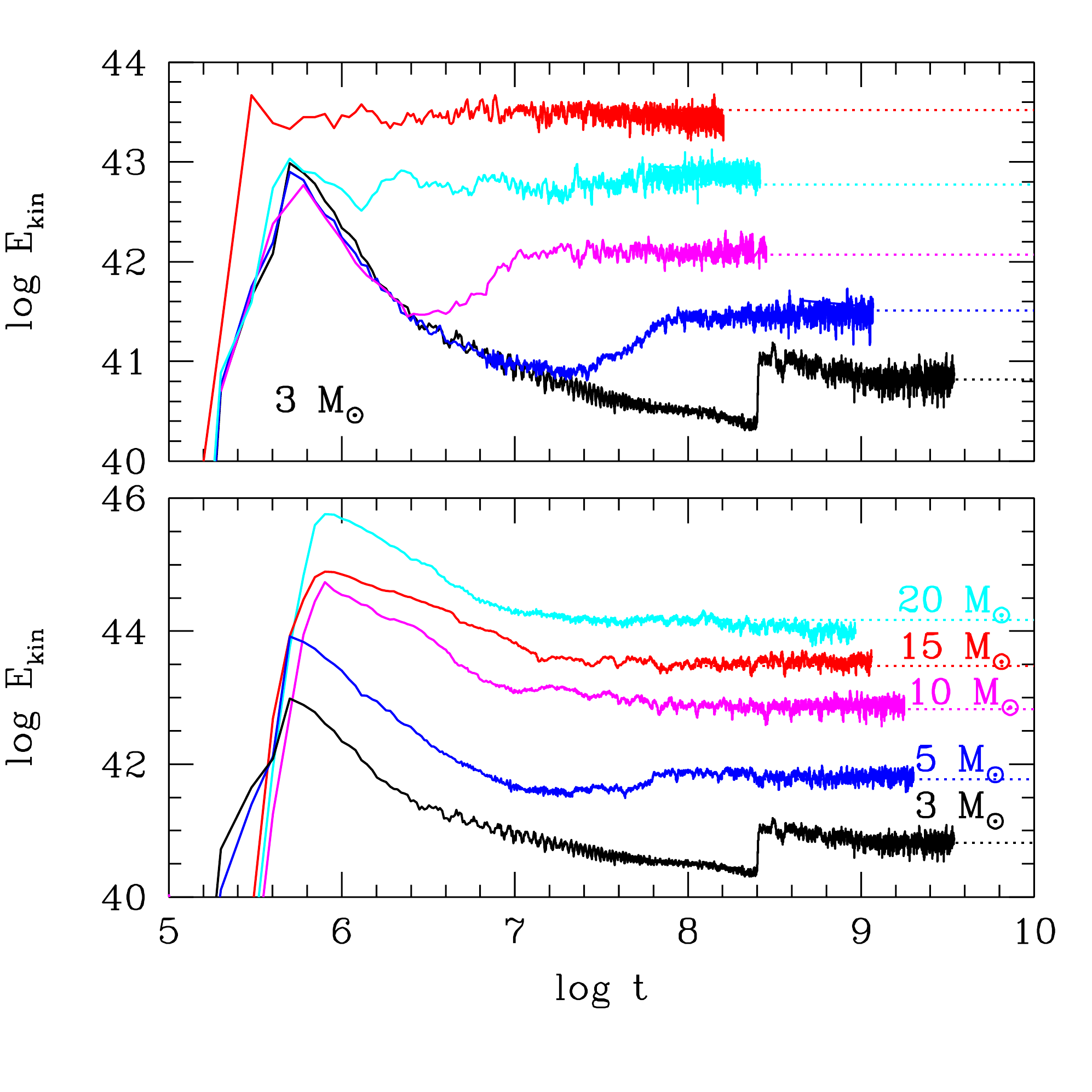}
   \caption{ Evolution of the total kinetic energy (in erg; y-axis with a base-10 log scale) as a function of time (in s) for the simulations described in Tab. 2. Top panel: results for 3 $\msun$ models with various luminosity enhancement factors:  3L0 (black), 3L1 (blue), 3L2 (magenta), 3L3 (cyan) and 3L4 (red). Bottom panel: results for a range of stellar masses.
   The dotted line for each model corresponds to the value of the total kinetic energy at the beginning of the steady state for convection.}
 \label{tekin_fig}
\end{figure}

\begin{figure}
\vspace{-2cm}
\includegraphics[height=16cm,width=8cm]{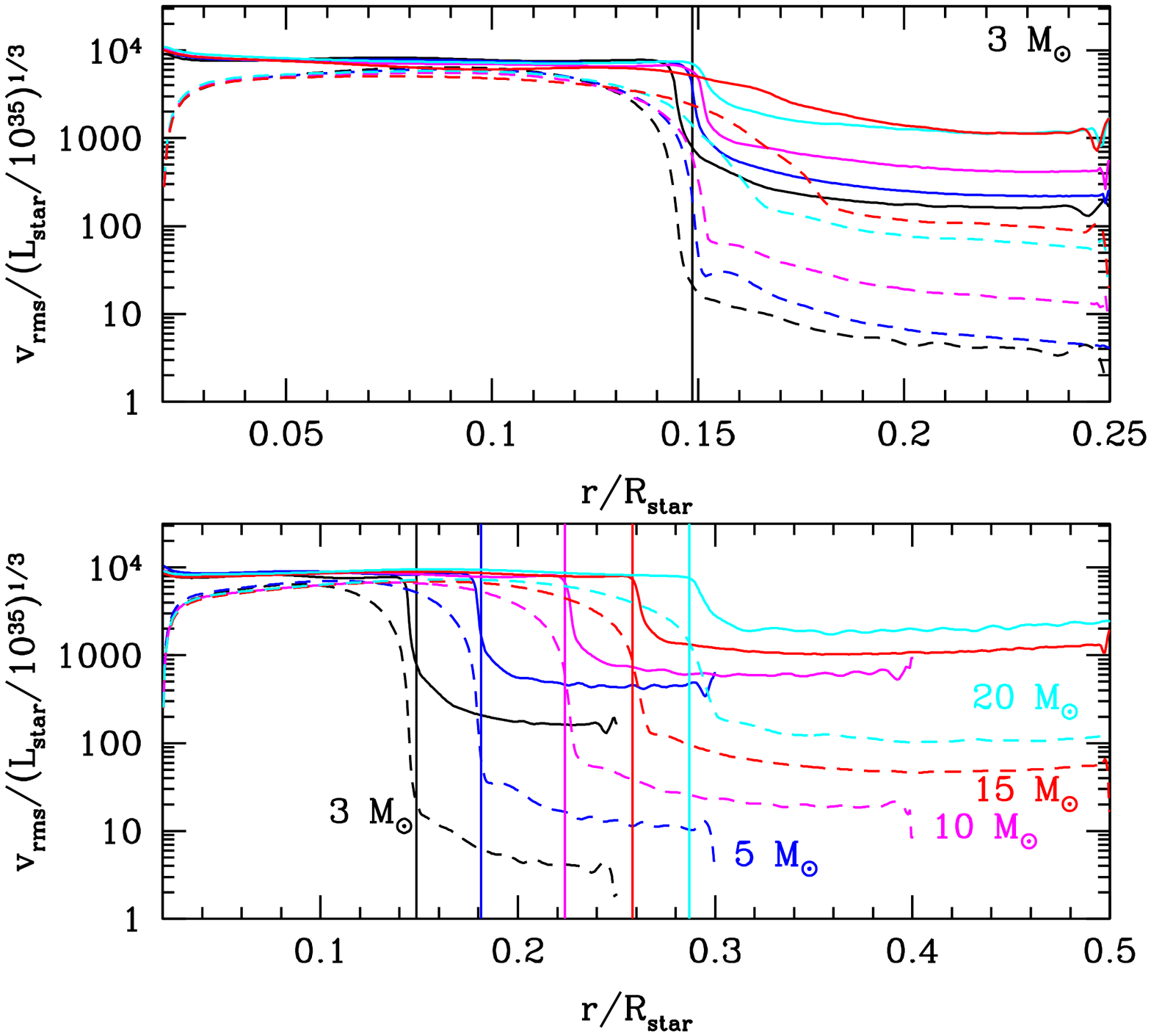}
\vspace{-2cm}
   \caption{ Radial profile of the time averaged rms velocity (solid lines) and rms radial velocity (dashed lines) scaled by $(L_{\rm star}/10^{35})^{1/3}$. 
   Top panel: results for 3 $\msun$ models with various luminosity enhancement factors:  3L0 (black), 3L1 (blue), 3L2 (magenta), 3L3 (cyan) and 3L4 (red). Bottom panel: results for a range of stellar masses: 3L0 (black), 5L0 (blue), 10L0 (magenta),15L0 (red) and 20L0 (cyan). The convective boundary corresponding to the Schwarzschild boundary from the 1D initial model is indicated by a vertical solid line with the colour corresponding to each stellar mass.}
 \label{vrms_fig}
\end{figure}

\section{Results: average dynamics}
\label{dynamics}

\begin{table*}
   \caption{Main properties of the 2D simulations.}
   \label{tab2}
   \begin{tabular}{lcccccccc} 
     \hline \hline
     Model &  $M/\msun$  & $L$ (erg/s) & $N_r \times N_\theta$ & $r_{\rm out}/R_{\rm star} $ & $\tau_{\rm conv}^{a}$ (s) &  $N_{\rm conv}^{b}$ &  $t_{\rm steady}^{c}$ (s) & $t_{\rm sim}^{d}$ (s)  \\
      \hline
       3L0 &  3 & 2.981 $\times  10^{35}$  & 336 x 168 & 0.25 & 1.9 $\times 10^6$ & 1442 & 9.5 $\times 10^8$ & 3.71 $\times 10^9$ \\ 
       3L1 & 3 & 2.981 $\times  10^{36}$  & 336 x 168  & 0.25 & 8 $\times 10^5$ & 1211 & 4.6 $\times 10^8$ & 1.43 $\times 10^9$ \\      
       3L2 &  3&  2.981 $\times  10^{37}$ & 336 x 168 & 0.25 & 3.9 $\times 10^5$ & 501 & 9 $\times 10^7$ & 2.84 $\times 10^8$ \\        
       3L2xhres &  3&  2.981 $\times  10^{37}$ & 684 x 342 & 0.25 & 3.8 $\times 10^5$ & 514 & 9 $\times 10^7$ & 2.84 $\times 10^8$ \\     
        3L3 &  3&  2.981 $\times  10^{38}$ & 336 x 168 & 0.25 & 1.7 $\times 10^5$ & 1904 & 6 $\times 10^7$ & 3.81$\times 10^8$ \\    
        3L3xhres &  3&  2.981 $\times  10^{38}$ & 684 x 342 & 0.25 & 1.7 $\times 10^5$ & 1243 & 6 $\times 10^7$ & 2.71 $\times 10^8$ \\             
       3L4 &  3& 2.981 $\times  10^{39}$  & 336 x 168 & 0.25 & 8.9 $\times 10^4$ & 1457 & 3 $\times 10^7$ & 1.60 $\times 10^8$ \\      
       3L4xhres &  3& 2.981 $\times  10^{39}$  & 684 x 342 & 0.25 & 8.7 $\times 10^4$ & 1400 & 3 $\times 10^7$ & 1.52 $\times 10^8$ \\          
       & & & & & & & & \\
        5L0 &  5 & 2.003 $\times  10^{36}$ & 400 x 200 &  0.3 & 1.4 $\times 10^6$ & 1260 & 2.45  $\times 10^8$ &   2.01 $\times 10^9$\\         
        10L0  &  10 & 2.139 $\times  10^{37}$ & 416 x 208 & 0.4 & 1.2 $\times 10^6$& 1260 & $2.1 \times 10^8$ & 1.77  $ \times 10^9$\\          
         15L0 & 15 & 7.387 $\times  10^{37}$ & 688 x 344 & 0.5 & 1.1  $\times 10^6$ & 875  &  $10^8$ &  1.14$ \times 10^9$\\          
          20L0 & 20 & 1.649 $\times  10^{38}$ & 864 x 430 & 0.6 & 1.1 $\times 10^6$ &800  &  9 $\times$ $10^7$ & 9.99 $ \times 10^8$\\   
      \hline

\multicolumn{9}{l}
{$^a$ Convective turnover time (see Sect. \ref{dynamics} for its definition).} \\
\multicolumn{9}{l}{$^b$ Number of convective turnover times covered by the simulation once  steady state convection is reached. }\\
\multicolumn{9}{l}{$^c$Physical time to reach a  steady state for convection.}\\
\multicolumn{9}{l}{$^d$Total physical runtime of the simulation.} \\
     \end{tabular}
    
\end{table*}

The properties of all simulations are summarised in Table \ref{tab2}. 
We define $t_{\rm steady}$ as the time required to reach a steady state for convection, characterised by the total kinetic energy $E_{\rm kin}$ of the system reaching a plateau. Before $t_{\rm steady}$, the initial relaxation phase is characterised by the propagation of strong acoustic waves and the onset of convection. At $t_{\rm steady}$, the value of the kinetic energy starts to stabilise and from this time it remains roughly constant with time (following the dotted curve which corresponds to the value of $E_{\rm kin}$ at $t_{\rm steady}$ for each model). The simulations are stopped at time $t_{\rm sim}$ provided in Table \ref{tab2}.
None of these simulations are thermally relaxed, given that the total simulation times for all models are orders of magnitude smaller than the relevant thermal timescale $\sim G M^2/(R_{\rm star}L$). As a consequence all these simulations are expected to maintain a secular drift. We have compared the radial profile  of the internal energy, averaged in the angular direction, for each 2D model at time $t_{\rm steady}$ and at time $t_{\rm sim}$.
We find a maximum of 0.5\% relative difference for the internal energy at a given radius, with the largest difference found for the most luminous models (see Sect. \ref{thermal}).
The above-mentioned drift  is thus so slow that calculating statistical or averaged data during this very slowly changing transitional state is sensible.

 Figure  \ref{tekin_fig} shows the evolution of the total kinetic energy as a function of time for all models and the plateau  characterising their steady state. The initial transient phase can last  a relatively long time, depending on the model studied. For the model 3L0,  
we note a different behaviour. After the peak due to strong acoustic waves, the kinetic energy continuously decreases until $t \sim 2.4 \times 10^8$ s (log $t \sim 8.38$). In this regime, convection develops in the core (within the 1D Schwarzschild boundary) in two spatially separate regions.
The abrupt increase of  $E_{\rm kin}$ observed at $t \sim 2.4 \times 10^8$ s marks the merging of these two convective regions and the beginning of fully developed convection in the core. The Mach number characterising the convective velocities in  model 3L0 is small, of the order of $\sim 10^{-4}$, which is
numerically challenging. This  low Mach number explains why several previous works artificially enhance the luminosity of the model \cp[][]{rogers13, horst20}.
There is no need for this artefact for the model 3L0 as MUSIC's numerical scheme allows convection to develop and eventually reach a steady state  even after a long transient phase.  Note that this unusual transient phase observed for the model 3L0 will likely change with a different procedure for initialising the simulation. All simulations start without an imposed background noise (i.e. initial velocities are set to zero). Imposing initially a background noise for the model 3L0 may change the location where convection starts and thus the behaviour of the transient phase, which is irrelevant for the  analysis performed in the following.
A global convective turnover time $\tau_{\rm conv}$ is estimated based on the rms velocity $ \vel_\mathsf{rms}(r,t)$ at radius $r$ and time $t$, which characterises a bulk convective velocity. We define $\tau_{\rm conv}$ by:

\begin{equation}
\tau_{\rm conv} =  \Big \langle {{ \int_{r_{\rm in}}^{r_{\rm conv}} {{\rm d}r \over \vel_\mathsf{rms}(r,t)  }  }  }\Big \rangle_t,
\end{equation}
where the rms velocity is given by
\begin{equation}
\vel_\mathsf{rms}(r,t) = \sqrt {\langle \velvec^2(r,\theta,t) \rangle_\theta}, \label{eq_rms}
\end{equation}
with $\velvec^2 = \velvec_r^2 + \velvec_{\theta}^2$, $\velvec_r$ and $\velvec_{\theta}$ being the radial and angular velocities, respectively.  
Time averages are denoted by $\langle \rangle_t$ and calculated between $t_{\rm steady}$ and $t_{\rm sim}$, the final time reached by the simulation (see values in Table \ref{tab2}). For any quantity $X$ we define:
\begin{equation}
\big \langle  X \big \rangle_t = {1 \over (t_{\rm sim} - t_{\rm steady})} \int_{t_{\rm steady}}^{t_{\rm sim}} X {\rm d}t
\end{equation} 
 The volume-weighted average in the angular direction $\langle \rangle_\theta$ is defined for any quantity X as:
\begin{equation}
\label{horizontal}
\big \langle X(r,\theta,t) \big \rangle_\theta = {\int_\theta  X(r,\theta,t) {\rm d}V(r,\theta) \over \int_\theta {\rm d}V(r,\theta)}.
\end{equation}
\noindent

The simulations are stopped after a time $t_{\rm sim}$ when convergence of the statistics used to determine the size of 
the layer penetrated by plumes is obtained, as explained in the next section (Sect.  \ref{statistics}). Table \ref{tab2} provides the values and numbers of the convective turnover times, respectively. Figure \ref{vrms_fig} displays the rms  velocity and rms radial velocity for  the 3 $\msun$ models with artificially enhanced luminosities (upper panel) and for the range of stellar masses investigated (lower panel), scaled by $L_{\rm star}^{1/3}$. In the convective core, our simulations reproduce the expected scaling of convective velocity  with luminosity $ \vel_\mathsf{conv} \propto L^{1/3}$ recovered by many hydrodynamical simulations \cp[e.g.][]{jones17, edelmann19, andrassy20, horst20, higl21, baraffe21}. This scaling is expected from mixing-length theory  based on the argument that the turbulent dissipation rate of kinetic energy in a turbulent convective zone scales with $\vel^3$ \cp{biermann32}. But a general scaling of the total flux with $\vel^3$ can also be derived for the kinetic energy and the enthalpy fluxes based on simple dimensional arguments \cp[see][]{jones17}
 
The rms velocities in the stably stratified region are due to the penetrative flows just above the convective boundary and to the propagation of internal waves excited by the convective motions and the penetrating plumes.  The top panel of Fig. \ref{vrms_fig}  shows that these velocities also increase with the luminosity,  suggesting more efficient overshooting of the convective motions above the convective boundary and thus larger overshooting length with increasing luminosity.
\ct{baraffe21} reports similar behaviours for convective envelopes of solar-like models with artificially enhanced luminosities.  Quantitative estimate of the overshooting lengths for all models is performed in Sect.  \ref{statistics}. 

\section{Results: extent of the overshooting region}
\label{statistics}

\subsection{Determination of overshooting lengths}
To determine an overshooting  length, we adopt the same approach as in \ct{baraffe21} and initially inspired by the findings of \ct{pratt17}. This approach is based on the  analysis of the depth of all convective plumes that penetrate beyond the convective boundary. The two criteria used to determine the depth of a penetrative plume at a given angle $\theta$  and time $t$ are based on the first zero above the convective boundary $r_{\rm conv}$ of the vertical kinetic energy flux $\mathbf f_{\rm k}$ and vertical heat flux $\mathbf f_{\rm \delta T}$, defined by \cp[see][]{pratt17}:
\begin{eqnarray}
 \mathbf f_{\rm k}(r,\theta,t) =  {1 \over 2} \rho(r,\theta,t) \velvec^2(r,\theta,t)  \velvec_r(r,\theta,t),  \label{eqk}\\
 \mathbf f_{\rm \delta T}(r,\theta,t)= \rho(r,\theta,t) c_P(r,\theta,t) \delta T(r,\theta,t) \velvec_r(r,\theta,t),  \label{eqt}
 \end{eqnarray}
 where $c_P$ is the specific heat at constant pressure and the temperature fluctuation $\delta T$ is defined by:
\begin{equation}
\delta T(r,\theta,t) = T(r,\theta,t) - \Big \langle {\big \langle T(r,\theta,t)\big \rangle_\theta } \Big \rangle_t.
\label{eqdelta}
\end{equation}

The  method is the same as the one developed in \ct{baraffe21}  for convective envelopes. At each time $t$, we calculate at each angle $\theta$ the radial positions $r_0(\theta,t)$ of a plume corresponding to the first zero of $\mathbf f_{\rm k}$ and $\mathbf f_{\rm \delta T}$, respectively, above the convective boundary $r_{\rm conv}$. The corresponding overshooting length $l_0$
with respect to  $r_{\rm conv}$ is defined by
\begin{equation}
l_0 (\theta,t) =  r_0(\theta,t) - r_{\rm conv}. \label{l0}
\end{equation}
Figure \ref{zero_fig} illustrates the angular structure of the overshooting layer at an arbitrary time for the 10 $\msol$ stellar model. 

\begin{figure}
\includegraphics[height=9cm,width=8cm]{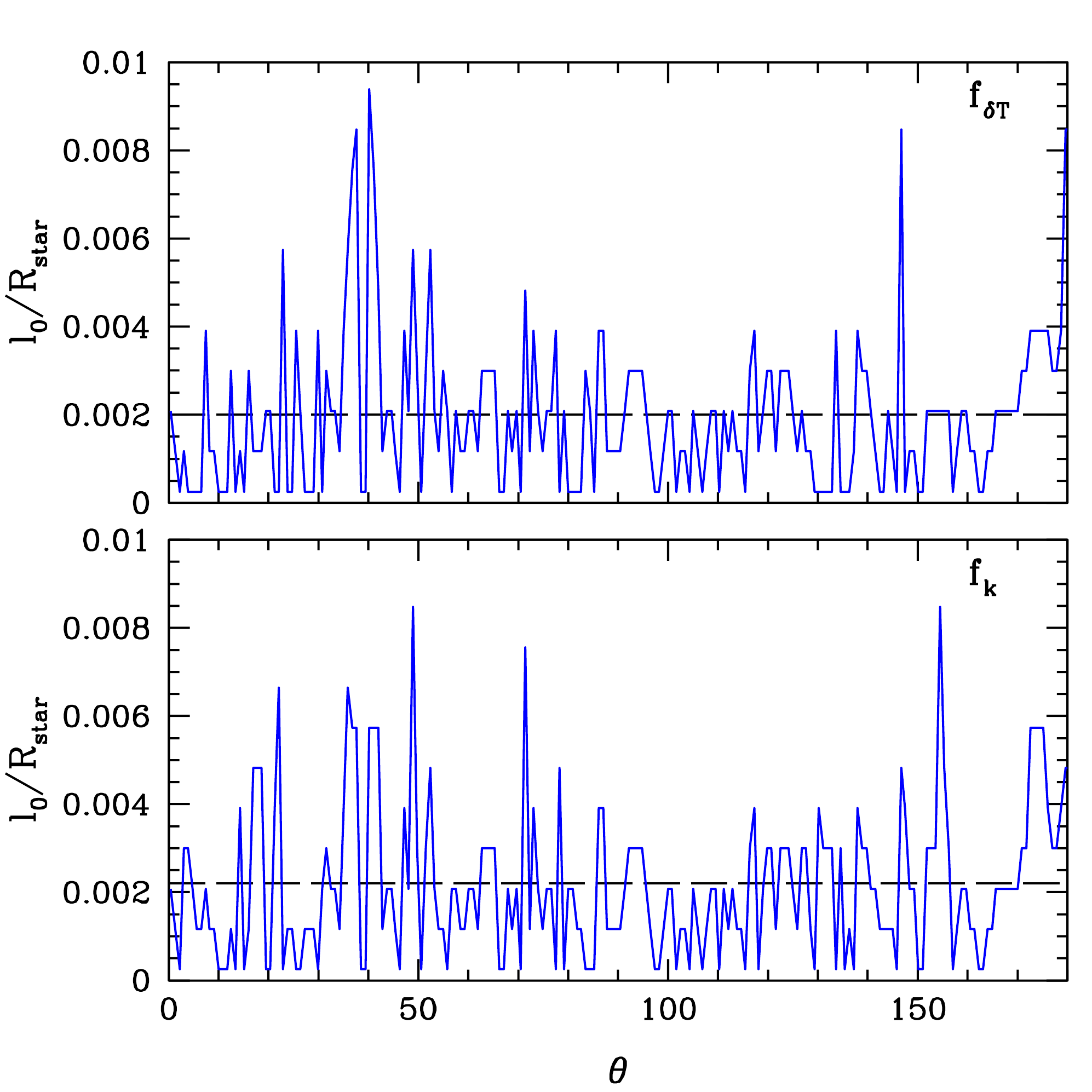} \label{zero_fig}
   \caption{Overshooting lengths  $l_0$ defined by Eq. (\ref{l0}) as a function of the angle $\theta$ at time $t=8.3 10^8 $s for the 10 $\msol$ model. The upper panel corresponds to $l_0$ defined by $\mathbf f_{\rm k}$ and the lower panel to $l_0$ defined by $\mathbf f_{\rm \delta T}$. The horizontal dashed line in each panel indicates the average overshooting length at this time.}
 \label{zero_fig}
\end{figure}

We then define the maximal overshooting length $l_0^{\rm max}$  at a given time by the maximum over all angles $\theta$:
\begin{equation}
 l_0^{\rm max}(t) = \mathsf{max}(l_0(\theta,t)). \label{lmax}
 \end{equation}

The time average  $l_{\rm max}$ = $\langle l_0^{\rm max}(t) \rangle_t$ provides an effective width for the overshooting layer where the most vigorous plumes penetrate and which we use to characterise the extension of the mixing layer over the long term evolution of the star \cp[][]{pratt17, baraffe21}.
Table \ref{tabdepth} displays  $l_{\rm max}$ based on the criterion for $\mathbf f_{\rm k}$ and $\mathbf f_{\rm \delta T}$, respectively, for  all models. 
The distributions of overshooting lengths derived from $\mathbf f_{\rm k}$ and $\mathbf f_{\rm \delta T}$, respectively, slowly converges with time, as found in \ct{pratt17} and \ct{baraffe21}. Several hundreds to thousand convective turnover times, depending on the stellar model, are required for the statistics to converge. Eventually, both criteria provide similar values for the effective overshooting width.
The values of the overshooting width based on $\mathbf f_{\rm \delta T}$ converge faster with time, compared to the value based on $\mathbf f_{\rm k}$, as found as well for convective envelopes in \ct{baraffe21}. The values of $l_{\rm max}(\mathbf f_{\rm \delta T})$ provided in  Table \ref{tabdepth} have reached a steady state for all models after $t_{\rm sim}$.  Depending on the stellar model,  $l_{\rm max}(\mathbf f_{\rm k})$ gets close to $l_{\rm max}(\mathbf f_{\rm \delta T})$ (difference of $\simlt$20\%) for all models but  models 3L0 and 20L0, for which $l_{\rm max}(\mathbf f_{\rm k})$ continues slowly decreasing even after more than 800 $\times \tau_{\rm conv}$. We run three simulations for the 3 $\msol$ models with enhanced luminosity with twice the resolution in both radial and angular directions and covering about the same simulation time as their lower resolution counterpart, in order to check the sensitivity of the values of $l_{\rm max}$ to the resolution. The properties of these higher resolution models (labelled 2xhres) are displayed in Table \ref{tab2}. The results for the overshooting lengths are given in Table \ref{tabdepth} and show similar values for $\mathbf {l_{\rm max}(\mathbf f_{\rm \delta T})}$  as found  with a lower resolution. The values for $l_{\rm max}(\mathbf f_{\rm k})$ of the higher resolution models are larger than the corresponding value for the lower resolution model, as it takes more time for $l_{\rm max}(\mathbf f_{\rm k})$ in the high resolution models to decrease to the level of $l_{\rm max}(\mathbf f_{\rm \delta T})$.  
But the value of $l_{\rm max}(\mathbf f_{\rm k})$ in the high resolution models continues decreasing with time and we expect it to eventually  converge and thus get much closer to $l_{\rm max}(\mathbf f_{\rm \delta T})$ and to the value of $l_{\rm max}(\mathbf f_{\rm k})$ found in the  lower resolution model.

\begin{table*}
   \caption{Effective width $l_{\rm max}$ of the overshooting layer  in units of the total stellar radius and of the pressure scale height at the convective boundary, for all models considered in this study. The quantity $l_{\rm max}(\mathbf f_{\rm k})$ is  based on the criterion using $\mathbf f_{\rm k}$ (Eq. \ref{eqk}) and $l_{\rm max}(\mathbf f_{\rm \delta T}$) is based on $\mathbf f_{\rm \delta T}$ (Eq. \ref{eqt}).}
   \label{tabdepth}
   \centering
   \begin{tabular}{l  c c c c } 
     \hline \hline
     Model &  $l_{\rm max}(\mathbf f_{\rm k})/R_{\rm star}$  &  $l_{\rm max}(\mathbf f_{\rm \delta T})/R_{\rm star}$   &  $l_{\rm max}(\mathbf f_{\rm k})/H_{P{\rm,CB}}$  &  $l_{\rm max}(\mathbf f_{\rm \delta T})/H_{P{\rm,CB}}$\\
      \hline 
      3L0 & 6.4 $\times 10^{-3}$ & 3.7  $\times 10^{-3}$ &  6.8 $\times10^{-2}$ &  3.9 $\times 10^{-2}$\\             
       3L1 & 4.2  $\times 10^{-3}$ & 4.2  $\times 10^{-3}$ &  $ 4.5 \times10^{-2}$ &  4.5 $\times 10^{-2}$ \\           %
        3L2 & 6.2  $\times 10^{-3}$&  6.1  $\times 10^{-3}$&  $ 6.6 \times10^{-2}$ &  6.5 $\times 10^{-2}$ \\           
        3L2xhres & 8.4  $\times 10^{-3}$&  6.4  $\times 10^{-3}$&  $  8.9\times10^{-2}$ & 6.8 $\times 10^{-2}$ \\    %
        3L3 & 1.8  $\times 10^{-2}$&  1.6  $\times 10^{-2}$& 1.9 $\times 10^{-1}$ & 1.7 $\times 10^{-1}$ \\   %
        3L3xhres & 2.2  $\times 10^{-2}$&  1.6  $\times 10^{-2}$& 2.3 $\times 10^{-1}$ & 1.7 $\times 10^{-1}$ \\ 
        3L4 & 3.5 $\times 10^{-2}$& 2.8  $\times 10^{-2}$ & 3.7 $\times 10^{-1}$ &3.0 $\times 10^{-1}$\\         %
        3L4xhres & 4.0 $\times 10^{-2}$& 3.0  $\times 10^{-2}$ & 4.2 $\times 10^{-1}$ & 3.2$\times 10^{-1}$\\ 
        
        & & & & \\
        
         5L0 & 9.3  $\times 10^{-3}$ & 6.0 $\times 10^{-3}$& 9.5 $\times 10^{-2}$& 6.1 $\times 10^{-2}$ \\    %
        10L0&1.2 $\times 10^{-2}$  &1.1 $\times 10^{-2}$ & 1.2 $\times 10^{-1}$&  1.1 $\times 10^{-1}$\\        %
        15L0 &  1.6 $\times 10^{-2}$ &1.3 $\times 10^{-2}$ & 1.66 $\times 10^{-1}$& 1.35 $\times 10^{-1}$\\     %
        20L0 &  3.5 $\times 10^{-2}$ &2.0 $\times 10^{-2}$ & 3.8 $\times 10^{-1}$& 2.17 $\times 10^{-1}$\\     %
      \hline
   \end{tabular}
\end{table*}

\subsection{Relationship between overshooting length and stellar luminosity}

\begin{figure}
\vspace{0cm}
\includegraphics[height=7cm,width=7cm]{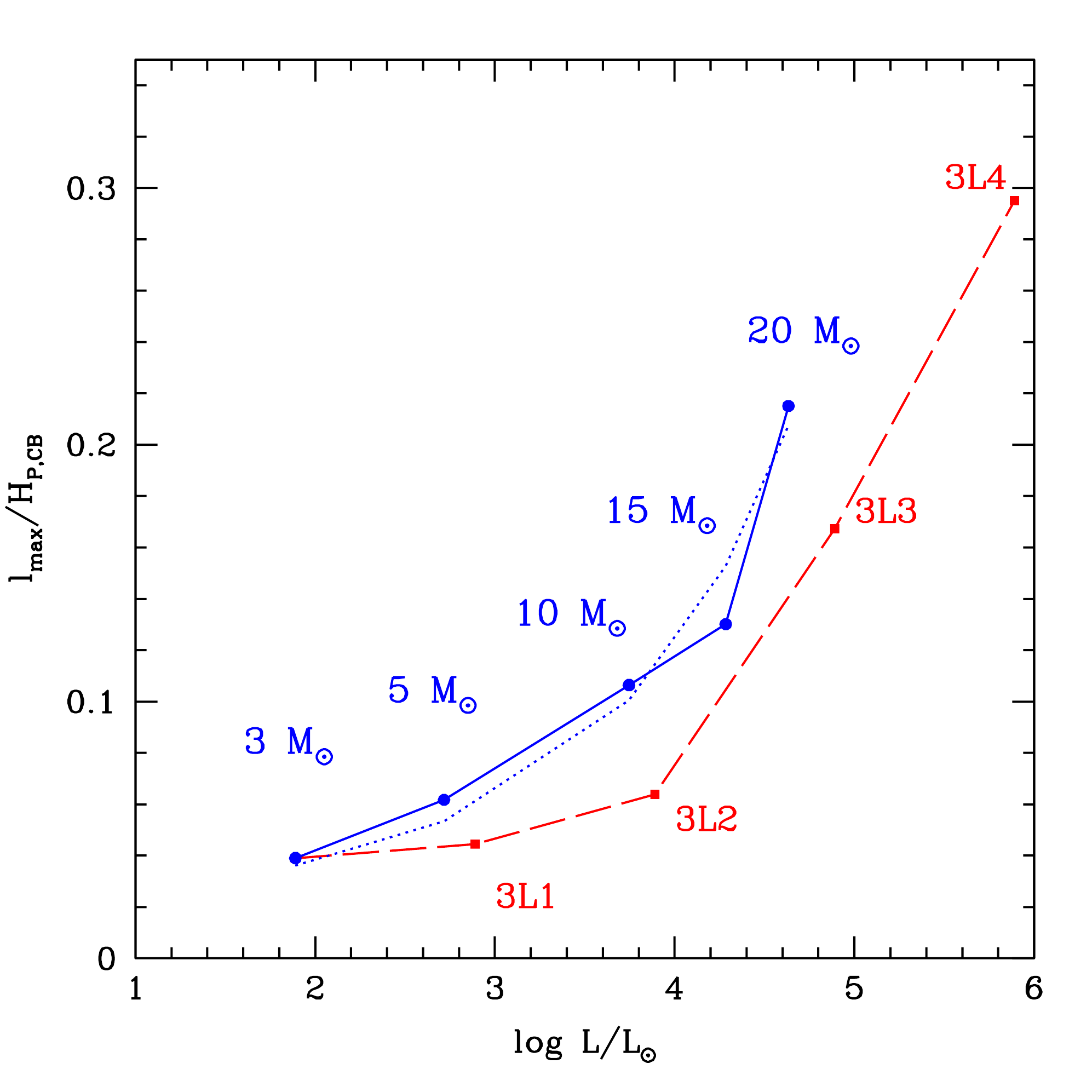}
\vspace{-0cm}
   \caption{Overshooting length $l_{\rm max}$,  in units of the pressure scale height at the convective boundary, as a function of the model luminosity. The 3 $\msol$ models with various luminosity enhancement factors are indicated in red (dashed line). The results for a range of stellar masses with realistic stellar luminosity are indicated in blue (solid line).  The dotted curve shows the fit for the overshooting length $d_{ov} / H_{\rm P{\rm,CB}}$ given by Eq. (\ref{scaling}).}
 \label{lov_fig}
\end{figure}

The variation of  $l_{\rm max}$ with the stellar luminosity is illustrated in Fig. \ref{lov_fig} for the 3$\msol$ models with enhanced luminosity and for the set of stellar masses with realistic luminosity. As expected from the behaviour of the rms velocities (see Fig. \ref{vrms_fig}) overshooting lengths increase with the stellar luminosity. 
To derive an approximate scaling relationship for the overshooting length $d_{\rm ov}$  that can be implemented in stellar evolution codes, we use the values of $l_{\rm max}$ derived from $\mathbf f_{\rm \delta T}$, since these values have converged with time. We derive the following expression which fits the results for the stellar mass range studied:
\begin{equation}
d_{\rm ov}/H_{\rm P{\rm,CB}} = 3.05 \times 10^{-3} \times (L/L_\odot)^{1/3} \times (r_{\rm conv}/H_{P{\rm,CB}})^{1/2} + 0.02 \label{scaling}
\end{equation}
 
 We find a typical scaling with the luminosity $d_{\rm ov} \propto L^{1/3} \propto \vel_{\rm conv}$. Numerical studies of convective envelopes report overshooting lengths $d_{\rm ov}$ which vary with the luminosity following $d_{\rm ov} \propto L^a$ with $a$ varying between 0.08 and 0.31 \cp[][]{hotta17, kapyla19, baraffe21}.  The analytical model of \ct[][]{zahn91} for penetration, based on first order estimate of the deceleration of a plume in an adiabatically stratified penetration zone,
 predicts $d_{\rm ov} \propto \vel_{\rm conv}^{3/2}$.
 Our results also show that the overshooting lengths derived for a fixed stellar mass (and thus a fixed convective core size) are systematically smaller than the one derived for larger cores but similar luminosity.
 Interestingly, a dependence of $d_{\rm ov}$ with the size of the core $r_{\rm conv}$ is also predicted by \ct[][]{zahn91} (see their Eq. (4.5)) with the same relation of proportionality $d_{\rm ov} \propto (r_{\rm conv}/H_{P{\rm,CB}})^{1/2}$ as found in  present simulations. This dependence in the Zahn model is derived from the strong variations with radius of various relevant quantities such as the gravitational acceleration $g$, the mass $m(r)$ enclosed in a sphere of radius $r$, the radiative conductivity $\chi$, and thus the radiative flux, close to the convective core boundary. In our simulations, we expect the radial dependence of the gravitational acceleration to have the main impact. We find that the larger the core (in terms of radius and mass), the smaller the gravitational acceleration at the core boundary  $g_{\rm conv} \sim G M_{\rm conv}/r_{\rm conv}^2$ (see values in Table \ref{tab1}). Therefore, the larger the stellar mass, the larger the velocities at the convective boundary and the smaller the restoring force due to gravity, implying up-flows to penetrate over larger distances. This is a plausible explanation  for the dependence of $d_{\rm ov}$  on the convective core radius.
 We  analyse below (Sect. \ref{1D}) whether the expression provided by Eq. (\ref{scaling}) provides a reasonable agreement between stellar evolution models and observations. 

\section{Thermal background evolution}
\label{thermal}

The  prescription used in the previous section to determine overshooting lengths  relies on two assumptions.  Firstly, we consider that the simulations have reached a steady state for convection ({\it i.e.} a global dynamical steady state). This assumption is reasonable  based on the observation that the total kinetic energy of the system reaches a plateau as a function of time (see Fig. \ref{tekin_fig}). Secondly, we assume that the relevant convective boundary from which the overshooting lengths are defined is the 1D Schwarzschild boundary. This is directly useful for the purpose of implementing these overshooting lengths in 1D stellar evolution codes. 
However, we find that  in all models a small nearly adiabatic layer just above the convective boundary forms rapidly once convection steady state is reached.   For the most luminous models, we observe that this small layer slowly grows in size with time.

\begin{figure}
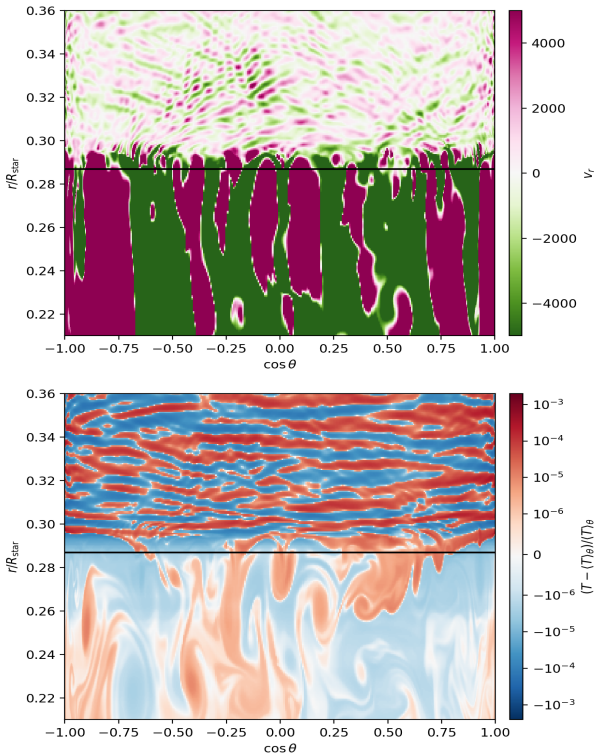

\includegraphics[height=5cm,width=8cm]{vel_MG.pdf}      
\includegraphics[height=5cm,width=8cm]{temp.pdf}.         

   \caption{Visualisation of the radial velocity $\vel_r$ [cm/s] (top panel) and the relative temperature fluctuations $(T - \langle T \rangle_\theta)/\langle T \rangle_\theta$  (bottom panel) in a region zoomed around the convective boundary (horizontal black line) for the model 20L0 at time $t=7 \times 10^8$ s. The x-axis represents the co-latitude (in terms of cos $\theta$). Note that to the better illustrate upwellings and downwellings in the top panel, the velocity scale is saturated, i.e. any velocity $> \vel_{\rm r,max}= 5\times 10^3$ cm/s ($ < \vel_{\rm r,min}= -5\times 10^3$ cm/s) are represented with the same color as  $\vel_{\rm r,max}$ ($\vel_{\rm r,min}$).}
 \label{visu_fig}
\end{figure}


 \ct[][]{anders22} also find a modification of the temperature gradient which becomes close to the adiabatic gradient in the penetration layer. They report that their simulations exhibit the process of convective {\it penetration} as defined by e.g. \ct{zahn91}, with convective penetrating motions mixing entropy  and establishing a nearly adiabatic stratification  above the Schwarzschild boundary \cp[see also][]{brummell02}.
  \ct[][]{anders22} suggest that the extent of convective penetration is limited and 
 derive arguments involving the convective flux, the viscous dissipation rate and the buoyancy work, providing an estimate of the penetration width. Depending on their setup, they find that penetration zones can take thousands of convective turnover times to saturate. They show properties of the flow and of the temperature fluctuations close to a convective boundary (see their Figure 1) which are similar to our results, as illustrated in Fig. \ref{visu_fig} for the model 20L0 at a given time.
As expected in convective regions,  convective upflows transport hot material from the central regions up to the top of the convective core. Inspection of  temperature fluctuations  ({\it i.e.} the difference between the local temperature and the horizontally averaged thermal background) indeed indicates that upflows in the convective region are characterised by positive temperature fluctuations and downflows by negative temperature fluctuations. When upflows cross the convective boundary, at the top of the convective core, and penetrate the stably stratified medium, they adiabatically expand and therefore get cooler (negative temperature fluctuation) and denser than the subadiabatically stratified environment.

To understand the establishment of a nearly adiabatic layer in the penetration region, one needs to compare  the advection timescale, which characterises the process of entropy mixing by penetrating flows ({\it i.e.} an advection process), and the thermal diffusion timescale.
If penetrating flows, as illustrated in the top panel of Fig. \ref{visu_fig}, can drive efficient entropy/thermal mixing,  the layer characterised by  penetrating up-flows will remain nearly adiabatic if thermal diffusion is slow enough. 
 Table \ref{tab_tau} provides  estimates of the diffusive timescale  $\tau_{\rm diff} = L^2/\kappa_{\rm rad}$ at the core boundary, with $L$ a relevant lengthscale and $\kappa_{\rm rad} = \chi/ (\rho c_P)$ the thermal diffusivity (which is the radiative diffusivity for present stellar models with $\chi$ defined in Eq. (\ref{eq:chirad})). Estimate of an advection timescale $\tau_{\rm adv} = L/\vel_{\rm r,rms}$ is based on the time averaged rms radial velocity at the core boundary.
For the characteristic lengthscales at the core boundary, we use  the overshooting distance $l_{\rm max}(\mathbf f_{\rm \delta T})$ (see Table \ref{tabdepth}) and the pressure scale height $H_{\rm P}$ (see Table \ref{tab1}). As illustrated in Table \ref{tab_tau}, typical advection timescales are much smaller than typical  thermal diffusion timescales for all models.

\begin{table*}
   \caption{Characteristic thermal diffusion timescales $\tau_{\rm diff} = L^2/\kappa_{\rm rad}$ and advection timescales $\tau_{\rm adv} = L/\vel_{\rm r,rms}$ (in s) estimated at the core boundary for all models, based on two  characteristic lentghscales,  $L=l_{\rm max}(\mathbf f_{\rm \delta T})$ and $L=H_{\rm P}$, respectively.
  $\kappa_{\rm rad}$ (in ${\rm cm^2 s^{-1}})$ is the thermal diffusivity  and $\vel_{\rm r,rms}$ (in cm ${\rm s}^{-1}$) is the time averaged rms radial velocity, both estimated at the core boundary. The last two columns provide the ratio ${\tau_{\rm diff} \over \tau_{\rm adv}} $ for the two lentghscales.}
   \label{tab_tau}
   \centering
   \begin{tabular}{l l l l l l l l l } 
     \hline \hline
     Model &  $\kappa_{\rm rad}$ & $\vel_{\rm r,rms}$ & $\tau_{\rm diff}(l_{\rm max})  $ & $\tau_{\rm diff}(H_{\rm P}) $  & $\tau_{\rm adv}(l_{\rm max})  $ & $\tau_{\rm adv}(H_{\rm P}) $ & ${\tau_{\rm diff} \over \tau_{\rm adv}} (l_{\rm max})$ & ${\tau_{\rm diff} \over \tau_{\rm adv}} (H_{\rm P}) $\\
      \hline
      3L0 & $10^7$ & 3.2$\times 10^{1}$ &2.6$\times 10^{10}$ & 1.7$\times 10^{13}$ & 1.6$\times 10^{7}$ & 4.1$\times 10^{8}$ & 1.6$\times 10^{3}$ & 4.1$\times 10^{4}$\\
      3L1 & $10^8$ & 7.8$\times 10^{2}$ & 3.3$\times 10^{9}$ & 1.7$\times 10^{12}$ & 7.4$\times 10^{5}$ & 1.7$\times 10^{7}$ & 4.4$\times 10^{3}$ & $ 10^{5}$ \\
       3L2 & $10^9$ &4.5$\times 10^{3}$ & 7$\times 10^{8}$ & 1.7$\times 10^{11}$ & 1.8$\times 10^{5}$ & 2.9$\times 10^{6}$ & 3.9$\times 10^{3}$ & 5.8$\times 10^{4}$ \\
      3L3 & $10^{10}$ & 2.1$\times 10^{4}$ &4.8$\times 10^{8}$ & 1.7$\times 10^{10}$ & $10^{5}$ & 6.2$\times 10^{5}$ & 4.8$\times 10^{3}$ & 2.7$\times 10^{4}$ \\
      3L4 & $10^{11}$ & 7.5$\times 10^{4}$ &1.5$\times 10^{8}$ & 1.7$\times 10^{9}$  & 5$\times 10^{4}$ &  1.8$\times 10^{5}$ &  3$\times 10^{3 }$ & $9.4\times 10^{3}$ \\ \\
      5L0 &  7$\times 10^7$ & 1.7 $\times 10^{2}$ &1.7$\times 10^{10}$ & 4.6$\times 10^{12}$  & 6.4$\times 10^{6}$ & $10^{8}$ & 2.6$\times 10^{3}$ & 4.6$\times 10^{4}$ \\
      10L0 & 7.5$\times 10^8$ &4.2$\times 10^{3}$ & 1.2$\times 10^{10}$ & 9.7 $10^{11}$ & 7$\times 10^{5}$ & 6.4$\times 10^{6}$ & 1.7$\times 10^{4}$ & 1.5$\times 10^{5}$ \\
      15L0 & 2$\times 10^9$ & 8.5$\times 10^{3}$ &$10^{10}$ & 5.4$\times 10^{11}$ & 5.2$\times 10^{5}$ & 3.9$\times 10^{6}$ & 1.9$\times 10^{4}$ & 1.4$\times 10^{5}$ \\
      20L0 & 4.5$\times 10^9$ &  1.6$\times 10^{4}$ & 1.4$\times 10^{10}$ & 3$\times 10^{11}$ & 5$\times 10^{5}$ & 2.3$\times 10^{6}$ & 2.8$\times 10^{4}$ & 1.3$\times 10^{5}$\\
 
      \hline
   \end{tabular}
\end{table*}

\begin{figure}
\vspace{-2cm}
\includegraphics[height=13cm,width=8cm]{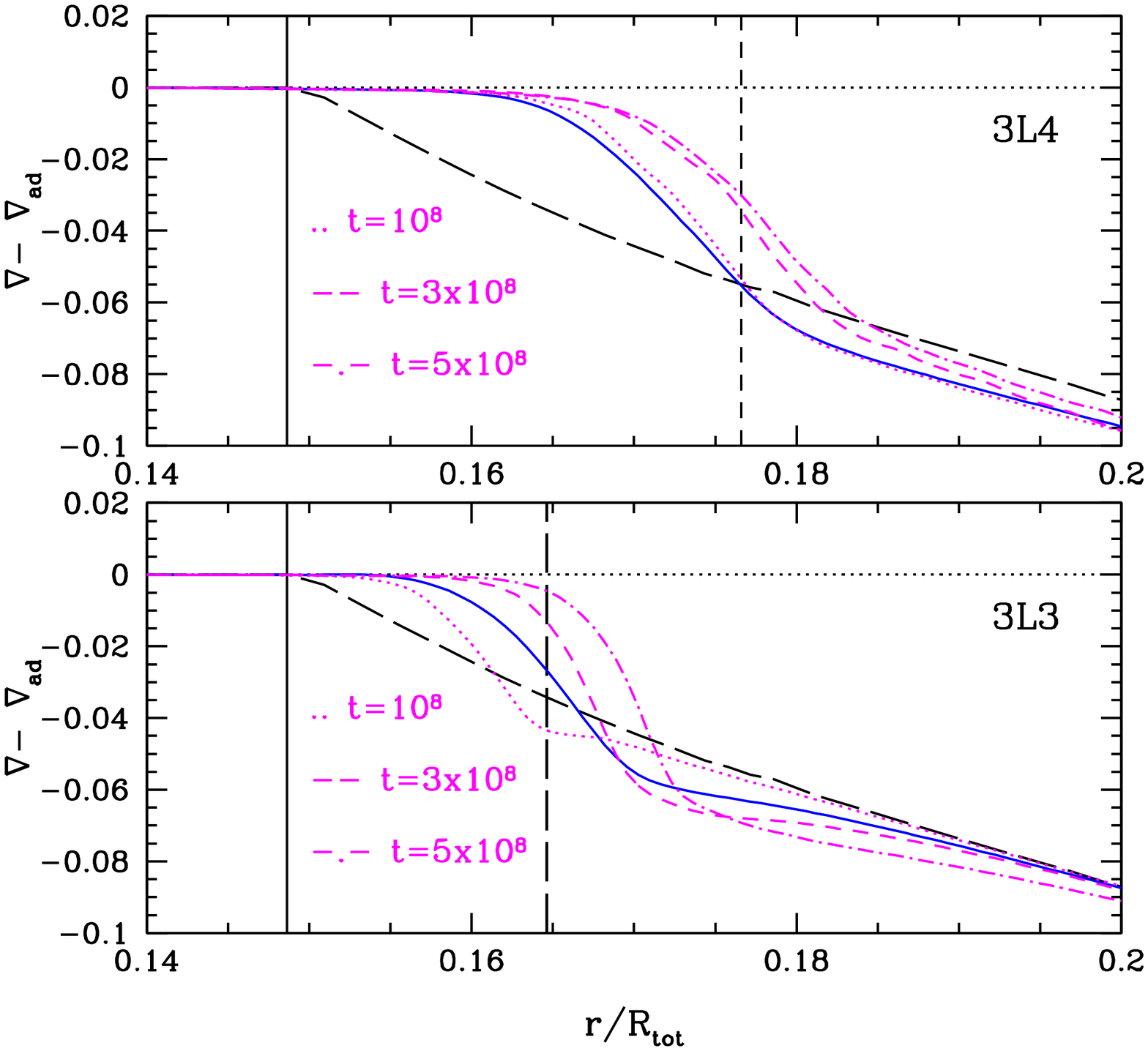}
\vspace{-2cm}
   \caption{Profile of the time and angular averages of the quantity ($\nabla-\nabla_{\rm ad}$) in the layers just above the convective core for the most luminous models. The 1D profile of ($\nabla-\nabla_{\rm ad}$)  is indicated by the black dashed line and the 1D convective core boundary by the vertical solid line. The location of $l_{\rm max}$ derived from $\mathbf f_{\rm \delta T}$ is indicated by the vertical dashed line. In both panels, the solid blue line corresponds to the time average between $t_{\rm steady}$ and $t_{\rm sim}$. The curves in magenta correspond to time averages over 20$\times \tau_{\rm conv}$ at a given time, as indicated in each panel (time $t$ in s).
}
 \label{delta_fig}
\end{figure}

The growth in size with time of the nearly adiabatic layer observed in the most luminous models is illustrated in Fig. \ref{delta_fig} for the models 3L3 and 3L4.  This growth with time may also happen in the less luminous models, but their very slow evolution and less vigorous penetrating flows may prevent clearly exhibiting this feature over present simulation times.
 We also note that the angular averaged temperature gradient in the models, while getting very close to the adiabatic gradient,  remains stable against the Schwarzschild criterion over the simulation times.

For the purpose of analysing the time evolution of the nearly adiabatic layer, we have extended the simulation time of the models 3L3 and 3L4 beyond the value of $t_{\rm sim}$ used to determine  overshooting depths (see Tab. \ref{tab2}), until $t_{\rm final}=5 \times 10^8$ s ($\sim 2600 \times \tau_{\rm conv}$ for 3L3 and $\sim 5300 \times \tau_{\rm conv}$ for 3L4). The aim is to reach a simulation time for these models close to or greater than the thermal diffusion timescale in the overshooting layer $\tau_{\rm diff}(l_{\rm max})$.
Given the smaller grid size and larger thermal diffusivity of these models, this is still computationally affordable. 
Figure \ref{delta_fig} shows clearly in models 3L3 and 3L4 that the radial extension of the nearly adiabatic layer slows down with time as the upper edge gets closer to the location of $l_{\rm max}$. Since the change of the temperature gradient is driven by penetrating flows, one may expect that the nearly adiabatic layer would not extend beyond $l_{\rm max}$ and that its growth may slow down when thermal diffusion starts to play a role. This process is likely to happen in the model 3L4, given that the final simulation time is significantly greater than the diffusive timescale over $l_{\rm max}$. It may start in the model 3L3 for which $t_{\rm final } \sim \tau_{\rm diff}(l_{\rm max})$. 

We cannot exclude that the modification of the temperature gradient is in part a transient effect in non-thermally relaxed simulations.
The initial conditions for the simulations are based on 1D stellar structures relying on the mixing-length theory (MLT) to describe the transport of heat in the convective core. A readjustment of the structure inducing a change of the location of the Schwarzschild boundary in the 2D simulations cannot be excluded, given the uncertainty inherent to the MLT. But the only process which can readjust the location of the convective boundary over present simulation times is the penetration of convective motions across the convective boundary.  A possible readjustment of the structure is thus also part of the process  that we aim at characterising in this work (see discussion in Sect. \ref{discussion}). 

\section{Application to 1D stellar evolution models and observations}
\label{1D}

\subsection{Spectroscopic Hertzsprung Russell diagram}
We  implement the scaling relationship  for the overshooting distance predicted by present simulations in stellar evolution models and compare these models to observations. 
For this purpose, the catalog of data of \ct{castro14} is relevant as it covers a large part of the stellar mass range investigated. This observational work provides the position of massive stars of spectral type OB in the Milky Way in the so-called spectroscopic Hertzsprung Russell diagram (sHRD). The sHRD uses a value {\calligra L }=$T_{\rm eff}^4/g$ in place of the stellar luminosity $L$, based on spectroscopically determined effective temperature $T_{\rm eff}$ and surface gravity $g$ of the star. The quantity {\calligra L } has the advantage that it can be calculated from stellar atmosphere analyses and compared to stellar evolution models without any knowledge of the distance or the extinction. \ct{castro14}
derived an empirical location in the sHRD of the zero-age-main-sequence (ZAMS), for stellar masses above $\sim 9 \msol$, and terminal-age-main-sequence (TAMS) that can be directly compared to stellar evolution tracks.  Because of the discrepancy in the main sequence width between observations and models, the main conclusion of their work is that convective core overshooting may be mass dependent and stronger than previously thought for stellar masses $\simgt 15 \msol$. We use this catalog of data to test the scaling relationship for $d_{\rm ov}$ predicted by present numerical simulations.

\begin{figure}
\hspace{-2cm}
\includegraphics[height=13cm,width=13cm]{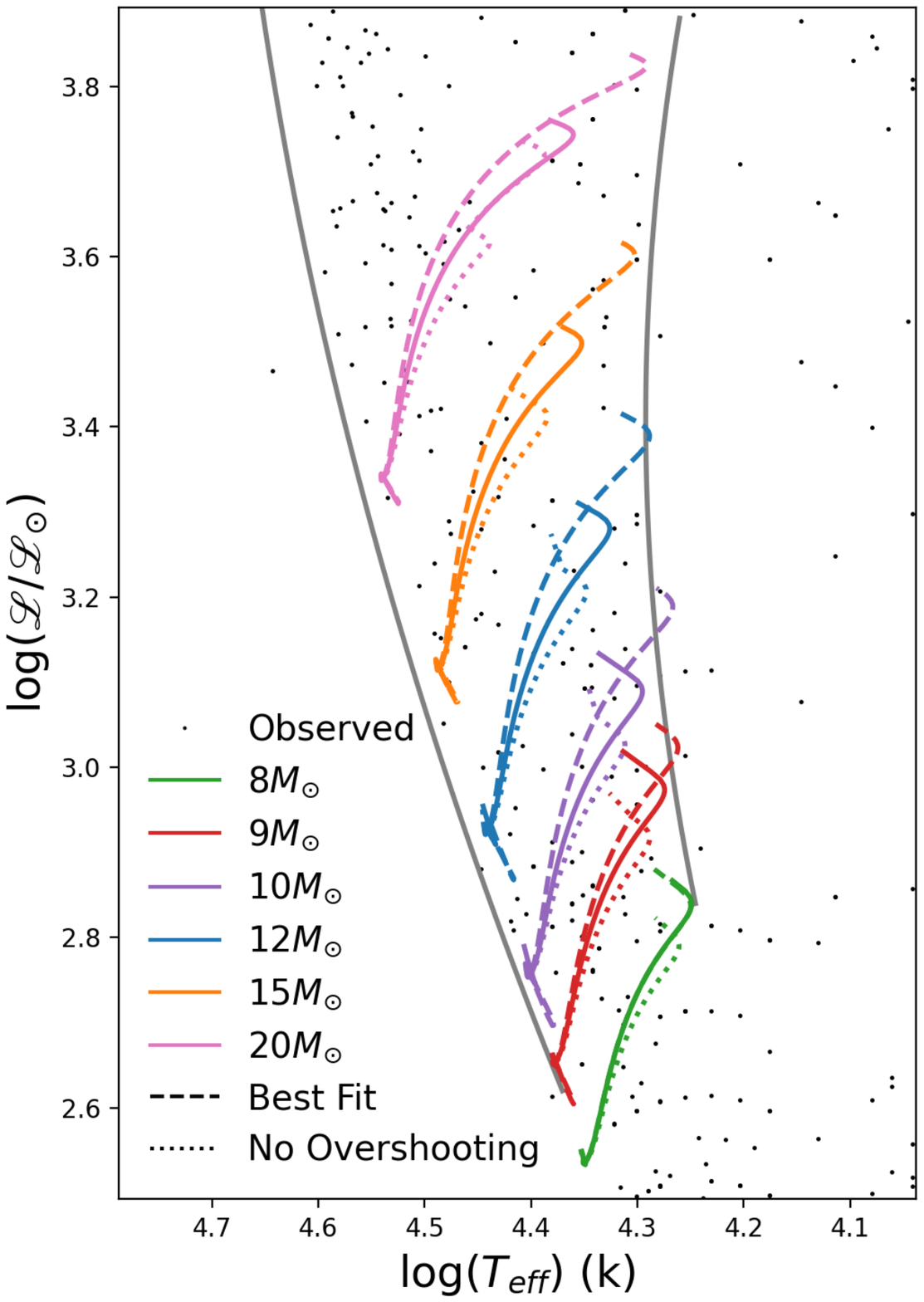}
\vspace{-2cm}
   \caption{Evolution of massive stars in the spectroscopic Herzsprung-Russell diagram with different treatments of core overshooting. The symbols are observed stars in the Milky Way from \ct{castro14}. The positions of the ZAMS and TAMS  are indicated by the black solid lines. Coloured solid lines: Models evolved with an  overshooting law given by Eq. (\ref{scaling}). Dashed lines: models evolved with an arbitrary overshooting length $d_{\rm ov}=\alpha_{\rm ov}  H_{\rm P}$ with values of $\alpha_{\rm ov}$ provided in Table \ref{tabdov}. Dotted lines: models with no overshooting.}
 \label{HRD}
\end{figure}

Stellar evolution models are calculated using the MESA code \cp{paxton11} which provides the flexibility of easily implementing the scaling relation for overshooting distance given by Eq. (\ref{scaling}). Instantaneous mixing is assumed over the distance $d_{\rm ov}$ above the convective core. We have performed calculations adopting either a radiative  or an adiabatic temperature gradient in the overshooting layer and find no significant impact on the evolutionary tracks. 
As done in \ct{castro14}, we compare the data to solar metallicity models. 
In Fig. \ref{HRD} we show the evolution of massive stars in the mass range 8-20 $\msol$ with no overshooting and with the scaling relationship given by Eq. (\ref{scaling}). The tracks are compared to the \ct{castro14} data and to the empirical locations of the ZAMS and the TAMS.  Table \ref{tabdov} provides the values of $d_{\rm ov}/ H_{\rm P}$ at the ZAMS and the TAMS, respectively, for the models evolved with the scaling relationship given by Eq. (\ref{scaling}).

\begin{table}
   \caption{Values of  $d_{\rm ov}/ H_{\rm P}$ for each stellar model evolved with the scaling relationship given by Eq. (\ref{scaling}) at the ZAMS and the TAMS, respectively. $\alpha_{\rm ov}$ is the fitted value for each stellar mass required to roughly reproduce  the observed main sequence width.}
   \label{tabdov}
   \centering
   \begin{tabular}{c c c c} 
     \hline \hline
     $M/\msol$ &   $d_{\rm ov}/ H_{\rm P}$ (ZAMS) & $d_{\rm ov}/ H_{\rm P}$ (TAMS) &  Fitted  $\alpha_{\rm ov}$  \\\\
      \hline
      8 & 0.09 & 0.11 & 0.1 \\ 
      9 & 0.10 & 0.13 & 0.2 \\
      10 & 0.11 & 0.14 & 0.3 \\
        12 & 0.13 & 0.18 & 0.35 \\
       15 & 0.16 & 0.23 & 0.4 \\
       20 & 0.22 & 0.32 & 0.45 \\
      \hline
   \end{tabular}
\end{table}

We have also computed models with an arbitrary overshooting length $d_{\rm ov}=\alpha_{\rm ov}  H_{\rm P}$ which is fixed for a given stellar mass but increases with mass. The values of $\alpha_{\rm ov}$ for this set of models are chosen to roughly reproduce the main sequence width and are provided in Table \ref{tabdov}.  We did not try to reproduce the ZAMS/TAMS empirical positions accurately. This set of models is also shown in Fig. \ref{HRD}.  
In agreement with the conclusions of \ct{castro14}, models without overshooting are unable to reproduce the observed main sequence width. An increasing overshooting distance with increasing stellar mass is required to reproduce the observed width. The overshooting scaling law based on our present hydrodynamical simulations predict this increase with the stellar mass (see Fig. \ref{lov_fig}).
It provides a good fit to the observed main sequence width for $M \simlt 10 \msol$.  But it tends to under-predict the value of  $d_{\rm ov}$ needed for models of higher mass to reach the observed location of the TAMS. A comparison of the values of $d_{\rm ov}$ given in Table \ref{tabdov} with the fitted values of $\alpha_{\rm ov}$ given in the same table suggests that values predicted by the hydrodynamical simulations are a factor $\sim$ 2 smaller than what is required to reach the observed location of the TAMS.


\subsection{Massive binaries}

We also test the overshooting scaling law given by Eq. (\ref{scaling}) against a selected sample of  massive eccentric binaries, namely  HD 152218 \cp{rauw16}, HD152219 \cp{rosu22a}  and CPD-41$^\circ$742 \cp{rosu22b}. We limit present analysis  to this restricted number of 
binary systems as they belong to the same young open cluster NGC 6231 and thus have the same metallicity, likely a solar metallicity. In addition,  their fundamental properties are inferred using the same methods and tools. This selected sample thus provides a small but homogeneous and consistent set of data to compare to stellar models. Their fundamental properties are provided in Table  \ref{binaries}.

\begin{figure*}
\vspace{-1cm}
\hspace{-4.3cm}
\includegraphics[height=15cm,width=10cm]{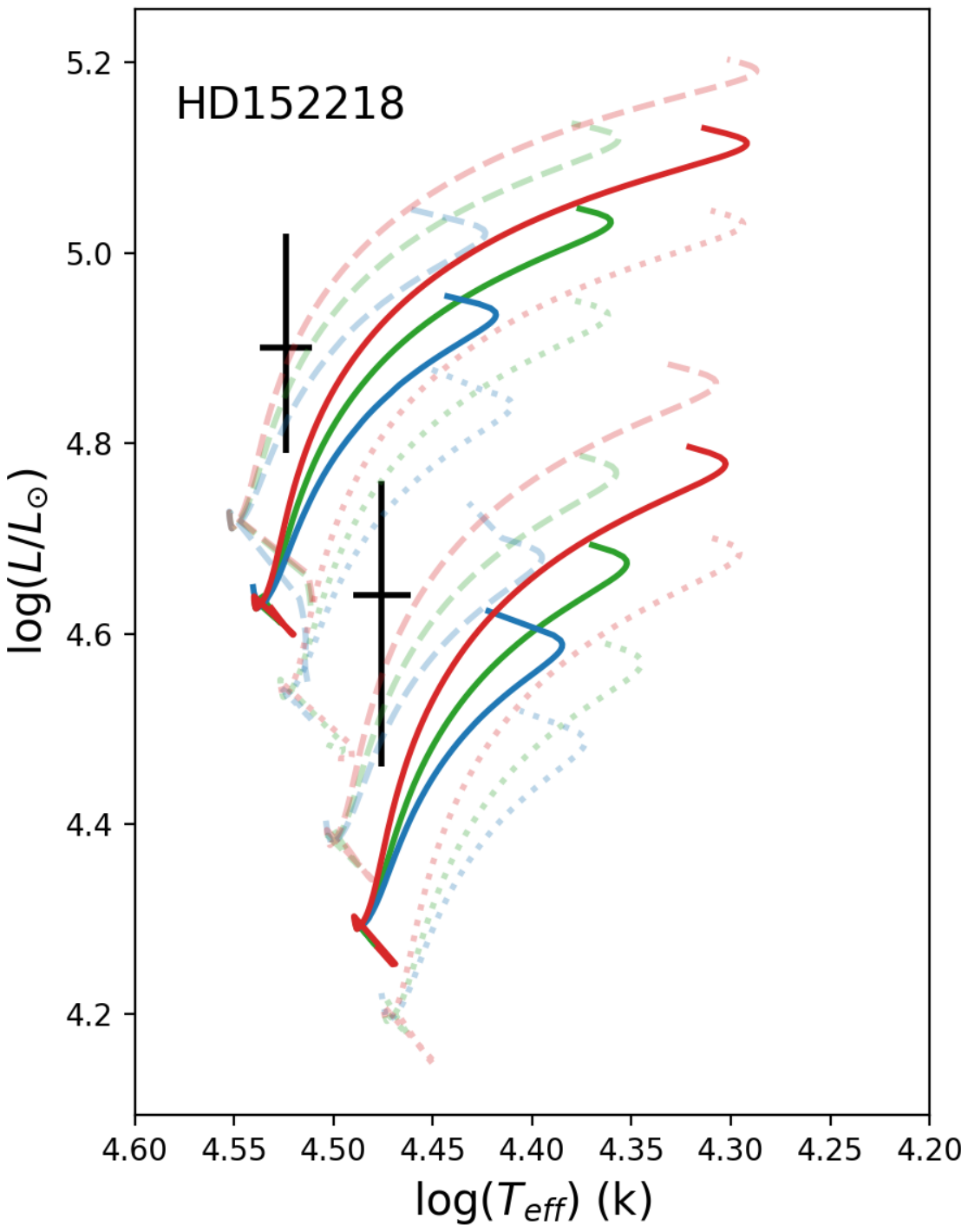}
\hspace{-4.3cm}
\includegraphics[height=15cm,width=10cm]{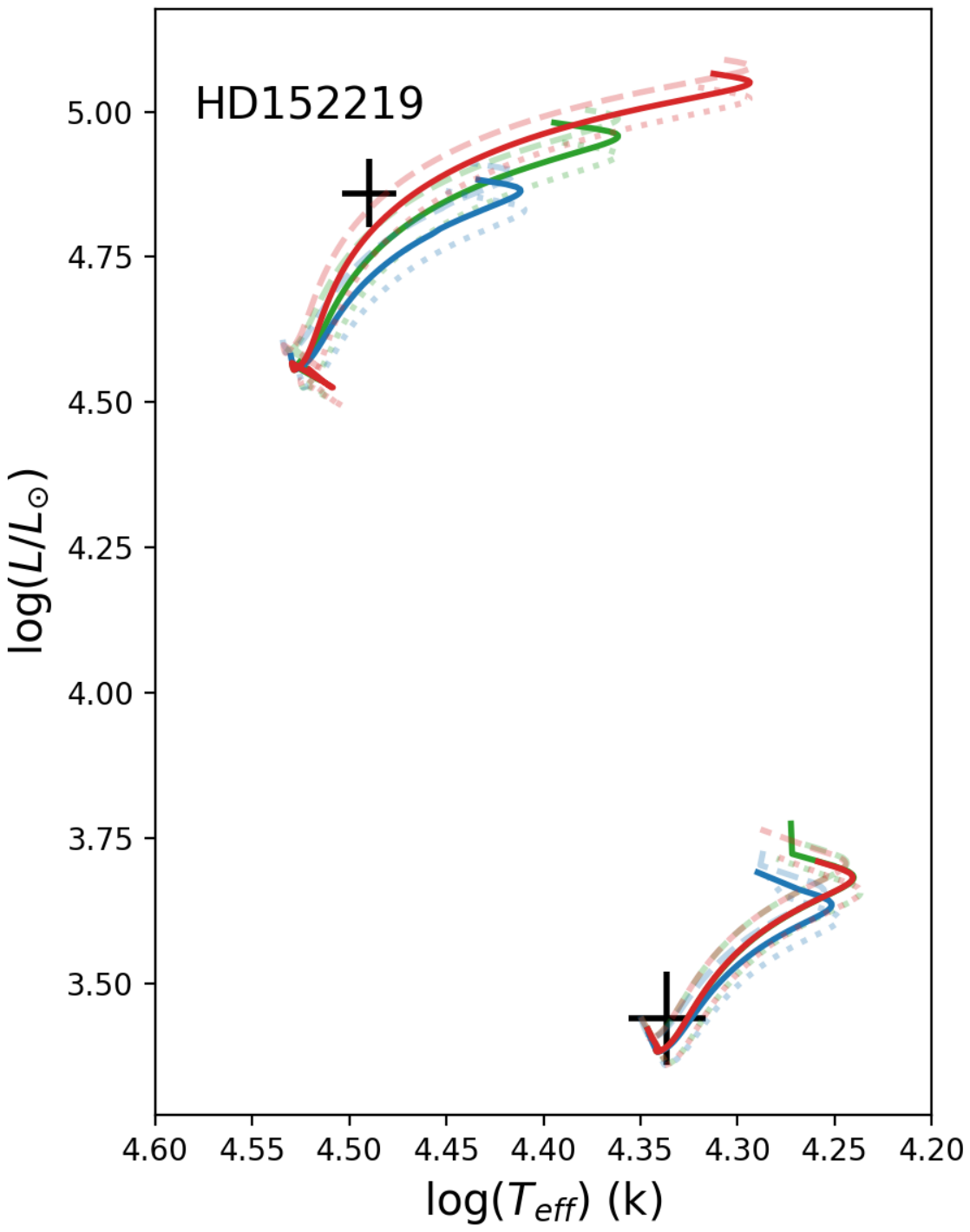}
\hspace{-4.3cm}
\includegraphics[height=15cm,width=10cm]{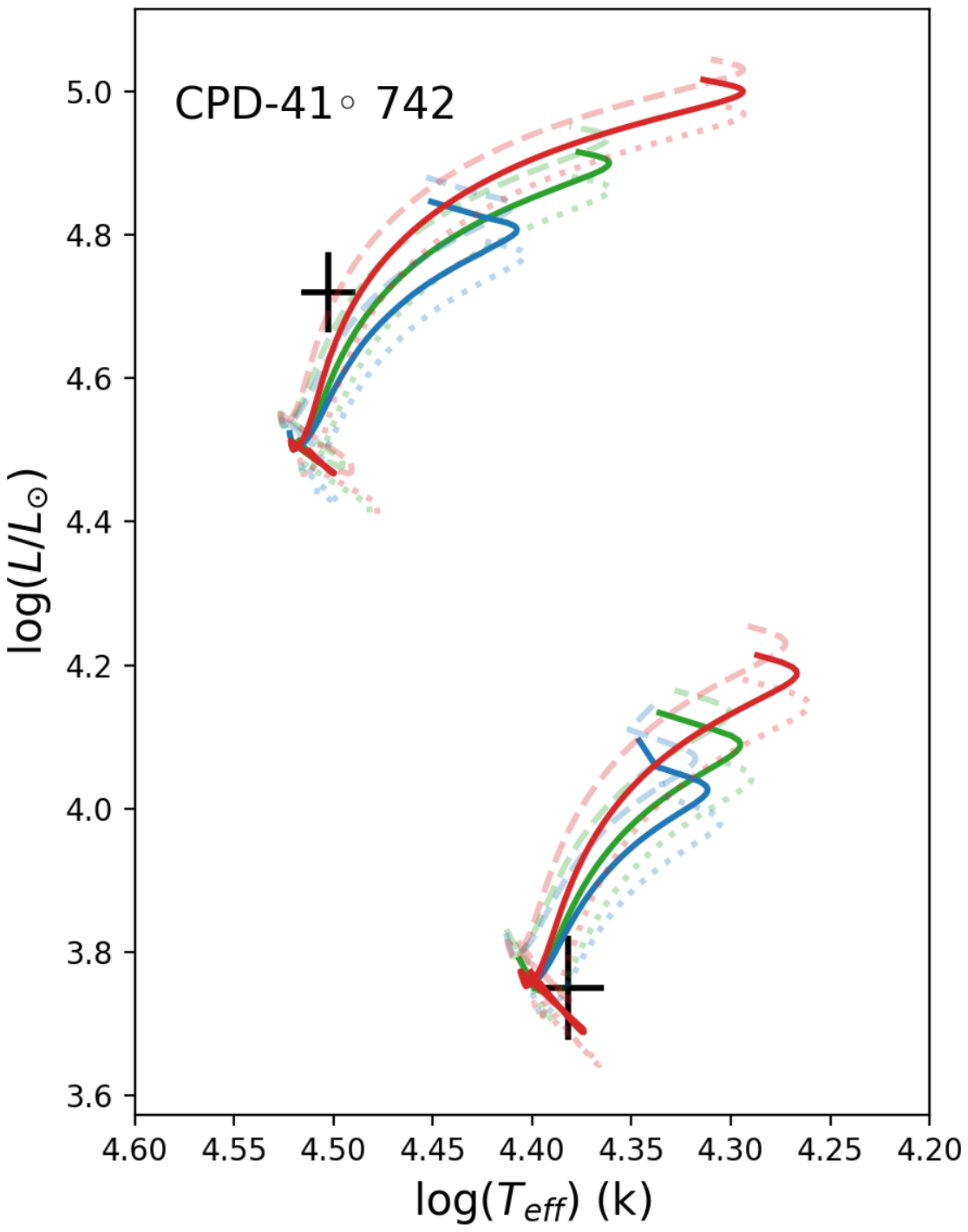}
\hspace{-4.3cm}
\vspace{-2cm}
   \caption{Comparison of evolutionary tracks with different treatments of overshooting and observations for massive binaries in the Hertzsprung-Russell diagram. Green lines: Models evolved with the  overshooting law given by Eq. (\ref{scaling}). Red lines: models evolved with an arbitrary overshooting length $d_{\rm ov}$ provided in Table \ref{tabdov} (Fittted $\alpha_{\rm ov}$). Blue lines: models without overshooting. The solid lines correspond to the track for the masses provided in Table \ref{binaries} and the dashed lines correspond to the tracks for the upper and lower masses within the errorbars. Observations are from \ct{rauw16} for HD 152218, \ct{rosu22a} for HD152219 and \ct{rosu22b} for CPD-41$^\circ$742.}
 \label{fig_binaries}
\end{figure*}

Figure \ref{fig_binaries} compares evolutionary tracks with different treatments of core overshooting with the observed properties of these binaries.
For HD 152218 (Fig. \ref{fig_binaries}, left panel), all models provide a solution within the error bars, but the large error bars do not provide a very stringent test for the treatment of overshooting. For HD 152219 (Fig. \ref{fig_binaries}, middle panel), all models provide a solution to the secondary, while only models with the arbitrary overshooting width from Tab. \ref{tabdov} provide a solution for the primary. Finally, for CPD-41$^\circ$742 (Fig. \ref{fig_binaries}, right panel), all models  fall within the error bars for the secondary. For the primary,  models with the arbitrary overshooting width provide a solution, but the models with the present hydrodynamical relationship provide solutions at the very limit of the error bars. Although this comparison of models with binaries is less conclusive than  the one performed with the \ct{castro14} data, it suggests that larger overshooting widths than predicted by the hydrodynamical relationship would provide a better fit, particularly for primaries with masses $\sim$ 18 $\msol$. 

\begin{table}
   \caption{Properties of the binaries used for the comparison with models.}
   \label{binaries}
   \centering
   \begin{tabular}{c c c c } 
     \hline 
     Binary & $M/\msol$ &  $T_{\rm eff}$(K) &  $L/L_\odot$ \\
      \hline
           & & & \\
      HD 152218a & 19.8 $\pm 1.5$ & 33 400 $\pm 1000$ & 7.94$^{+2.52}_{-1.77}  \times 10^4$ \\   
      HD 152218b & 15.0 $\pm 1.1$ & 29 900 $\pm 1000$ &  4.36$^{+1.39}_{-1.48}  \times 10^4$ \\    
      & & & \\
      HD 152219a & 18.64 $\pm 0.47$ & 30 900 $\pm 1000$ & (7.26$\pm 0.97) \times 10^4$ \\
      HD 152219b & 7.70 $\pm 0.12$ & 21 697 $\pm 1000$ & (2.73$\pm 0.51) \times 10^3$ \\
      & & & \\
      CPD-41$^\circ$742a  & 17.8 $\pm 0.5$ & 31 800 $\pm 1000$ & 5.28$^{+0.67}_{-0.68} \times 10^4$ \\
      CPD-41$^\circ$742b & 10.0 $\pm 0.3$ & 24 098 $\pm 1000$ & 5.58$^{+0.93}_{-0.94} \times 10^3$ \\
       \hline
   \end{tabular}
\end{table}

\section{Discussion and Conclusion}
\label{discussion}
This work is an initial, exploratory investigation, in which we infer an overshooting width $d_{\rm ov}$ for a  broad range of ZAMS stellar models based on hydrodynamical simulations.  
The present determination of an effective overshooting width, characterising the extent of mixing on the long term evolution of the star, is based on an approach relying on extreme events of penetrating flows  previously developed for convective envelopes of solar-type stars \cp[e.g.][]{pratt17, pratt20, baraffe21}. For ZAMS stars, we find that the overshooting distance scales with the stellar luminosity and the convective core radius, resulting in values of $d_{\rm ov}$ which significantly increase with stellar mass. Obtaining this increase is an important achievement,  since such an increase is suggested by several observational constraints.  But although the results within our framework are qualitatively in agreement with the observed trends, quantitatively, they are unable to match the available data. Indeed,  the comparison of stellar evolution tracks to the properties of a sample of Milky Way main sequence stars suggests that the predicted values of $d_{\rm ov}$ are underestimated for $M \simgt 10 \msol$. The comparison to massive binaries suggests the same limitation. This points to a need for further computational studies, as discussed below.

The diagnostics we have used to examine the present set of 2D simulations have their limitations and several physical or numerical ingredients may increase the values of overshooting lengths. One limitation is our assumption that the overshooting lengths determined within the extreme event framework, which is based  on fluxes in an Eulerian
approach, characterise the extension of efficient chemical mixing above the convective core. Quantifying the extent of chemical mixing is the prime interest for an application to 1D  stellar evolution models. This assumption can be verified with an analysis of mixing based on Lagrangian tracer particles, a direction that will be explored in a future work. The formation of a small nearly adiabatic layer above the convective core of our models due to efficient entropy mixing by the upward penetrating flows  indicates that  efficient chemical mixing should also proceed between the convective boundary and the location of the maximal overshooting length $l_{\rm max}$. But the size of the layer for efficient chemical mixing and the one of the nearly adiabatic layer are not expected to be the same, even if the same initial process drives thermal and chemical mixing ({\it i.e.}  advection by upward flows). Our results suggest that  the extent of the nearly adiabatic layer may be limited by thermal diffusion, as observed for the most luminous models when the simulation time exceeds the typical thermal diffusive timescale in the overshooting layer. Thermal diffusion will not limit the extent of chemical mixing. Internal waves excited by convective plumes at the core boundary could however  contribute to additional chemical mixing and extend the size of the chemical mixing layer beyond  $l_{\rm max}$. This is also under further investigation and could provide an interesting process to increase the overshooting lengths derived with present approach.

 Extension  to three-dimensional geometry is an obvious next step, since  the structure and the geometry of penetrating convective flows are expected to be modified in 3D compared to 2D simulations \cp[see][]{brummell02}.  Despite 2D convective velocities being  on average larger than 3D velocities \cp[][] {meakin07, pratt20}, 
 several works have suggested that the filling factor and plume geometry could be smaller in 3D than in 2D \cp[see discussion in][] {rogers06}. Simulations in 3D may thus provide larger  overshooting lengths, as needed to reproduce stellar observations. But so far no conclusive study of the filling factor and plume shape using the same simulation framework in 2D and 3D has been performed \cp[see][] {pratt20}.

 Further numerical studies need to be performed in order to determine the impact of rotation and whether it can provide another driver to increase overshooting lengths
 and/or to make mixing more efficient \cp[see e.g.][]{browning04}. A limitation of the present simulations, and indeed many global simulations of stars, is the fact that they are not thermally relaxed, since this would require simulation times even greater than the values for the thermal diffusion timescale over a pressure scale height $\tau_{\rm diff}(H_{\rm P})$  provided in Table \ref{tab_tau}. The direct
application of the overshooting lengths predicted by these simulations to ``real" stars must thus be taken with caution, since the final relaxed state for these simulations may have different properties from present non thermally relaxed states. This does not however preclude analysing the efficiency of overshooting as a function of stellar mass and luminosity  during the slowly evolving transient phase during which convection is considered to be in steady state. One can speculate that even if the convective boundary moves with respect to the initial 1D Schwarzschild boundary after thermal relaxation, the overshooting lengths determined on a dynamical steady state from this new boundary may still be close to the the ones determined in this work. Unfortunately, to verify this implies running the simulations over a thermal timescale, which  is computationally not feasible. More extreme enhancement factors for the luminosity could allow reaching  thermal relaxation. But as shown recently for convective envelopes in \ct[][]{baraffe21}, large enhancement factors can push the simulated conditions away from the original target star, inducing a significant drift from the initial stellar structure.

In addition, the scaling presented in this work  is derived for ZAMS stars and may not apply to cores that have evolved on the main sequence. Indeed, the development of a molecular weight gradient at the core boundary due to hydrogen burning  will most likely limit the radial penetration of upward flows above  the convective core boundary. Numerical simulations of the convective core of  main sequence 5 $\msol$ and 20 $\msol$ star models indicate much smaller values of $l_{\rm max}$ compared to their ZAMS counterpart (Morison et al., in prep). In addition, they  show no sign of entrainment which could result in an increase of the size of the convective core.  Whether 3D, rotation and/or other instabilities can solve the problem of ``impenetrability" of convective flows due to the building of a molecular weight gradient during the evolution on the main sequence is an open question.
Other effects and/or improvement of present 2D simulations are needed to increase the overshooting lengths for both ZAMS and main sequence models.

In conclusion, this work provides results which qualitatively validate the increase of  overshooting lengths with stellar mass (or stellar luminosity) suggested by observations \cp[e.g.][]{castro14}. Quantitatively, however, the predicted values are underestimated for stellar masses $\simgt 10 \msol$. Our present results apply only  to stellar models on the ZAMS. Our study illustrates the challenges and the promise of hydrodynamical simulations. It sets the stage for broader, and more physically detailed studies to resolve in the future this quantitative discrepancy with observations.

\section*{Acknowledgements}
This work is supported by the ERC grant No. 787361-COBOM and the consolidated STFC grant ST/R000395/1. We are grateful to Noberto Castro for providing data in a user friendly form and for useful advises for using the catalog. We thank our anonymous referee for very valuable comments and suggestions. The authors would like to acknowledge the use of the University of Exeter High-Performance Computing (HPC) facility ISCA and of the DiRAC Data Intensive service at Leicester, operated by the University of Leicester IT Services, which forms part of the STFC DiRAC HPC Facility. The equipment was funded by BEIS capital funding via STFC capital grants ST/K000373/1 and ST/R002363/1 and STFC DiRAC Operations grant ST/R001014/1. DiRAC is part of the National e-Infrastructure. Part of this work was performed under the auspices of the U.S. Department of Energy by Lawrence Livermore National Laboratory under Contract DE-AC52-07NA27344.

\section*{Data Availability}

The  1D initial structures are available on the repository: 

\noindent
http://perso.ens-lyon.fr/isabelle.baraffe/2Dcore$\_$overshooting$\_$2023. The other data underlying this article will be shared on reasonable request to the corresponding author.  

\bibliographystyle{mnras}
\bibliography{references}

\bsp	
\label{lastpage}
\end{document}